\documentclass[twocolumn,nofootinbib,english,superscriptaddress,amsmath,amssymb,floats,prx]{revtex4-1}
\usepackage[latin9]{inputenc}
\usepackage{color}
\usepackage{mathtools}
\usepackage{amsmath}
\usepackage{graphicx}

\makeatletter

\usepackage{comment}
\newcommand{\beq}{\begin{equation}}
\newcommand{\eeq}{\end{equation}}
\newcommand{\bea}{\begin{eqnarray}}
\newcommand{\eea}{\end{eqnarray}}
\newcommand{\br}{\mathbf{r}}

\usepackage{multibib}
\newcites{supp}{Supplementary}

\usepackage[bottom]{footmisc}

\@ifundefined{textcolor}{}{%
 \definecolor{BLACK}{gray}{0}
 \definecolor{WHITE}{gray}{1}
 \definecolor{RED}{rgb}{1,0,0}
 \definecolor{GREEN}{rgb}{0,1,0}
 \definecolor{BLUE}{rgb}{0,0,1}
 \definecolor{CYAN}{cmyk}{1,0,0,0}
 \definecolor{MAGENTA}{cmyk}{0,1,0,0}
 \definecolor{YELLOW}{cmyk}{0,0,1,0}
}

\usepackage{epstopdf}
\usepackage{array}
\usepackage{mathrsfs}
\usepackage{dcolumn}
\usepackage{bm}

\newcolumntype{L}[1]{>{\raggedright\let\newline\\\arraybackslash\hspace{0pt}}m{#1}}
\newcolumntype{C}[1]{>{\centering\let\newline\\\arraybackslash\hspace{0pt}}m{#1}}
\newcolumntype{R}[1]{>{\raggedleft\let\newline\\\arraybackslash\hspace{0pt}}m{#1}}

\graphicspath{{../Inkscape/},{../Mathematica/},{supplementary/}}

\newcommand{\mbf}[1]{\mathbf{#1}}

\usepackage{tikz}
\usetikzlibrary{calc,intersections}
\usetikzlibrary{shadings}
\usepgflibrary{shadings}

\usepackage{babel}

\makeatother

\begin{document}

\title{Modeling unconventional superconductivity at the crossover between strong and weak electronic interactions}

\author{Morten H. Christensen}
\thanks{These authors contributed equally to this work}
\affiliation{School of Physics and Astronomy, University of Minnesota, Minneapolis, MN 55455, USA}

\author{Xiaoyu Wang}
\thanks{These authors contributed equally to this work}
\affiliation{National High Magnetic Field Laboratory, Tallahassee, FL 32310, USA}

\author{Yoni Schattner}
\affiliation{Department of Physics, Stanford University, Stanford, CA 94305, USA}
\affiliation{Stanford Institute for Materials and Energy Sciences, SLAC National Accelerator Laboratory and Stanford University, Menlo Park, CA 94025, USA}

\author{Erez Berg}
\affiliation{Department of Condensed Matter Physics, Weizmann Institute of Science, Rehovot 76100, Israel}

\author{Rafael M. Fernandes}
\email{rfernand@umn.edu}
\affiliation{School of Physics and Astronomy, University of Minnesota, Minneapolis, MN 55455, USA}

\date{\today}
\begin{abstract}
High-temperature superconductivity emerges in many different quantum materials, often in regions of the phase diagram where the electronic kinetic energy is comparable with the electron-electron repulsion. Describing such intermediate-coupling regimes has proven challenging as standard perturbative approaches are inapplicable. Here, we employ Quantum Monte Carlo (QMC) methods to solve a multi-band Hubbard model that does not suffer from the sign-problem and in which only repulsive interband interactions are present. In contrast to previous sign-problem-free studies, we treat magnetic, superconducting and charge degrees of freedom on an equal footing. We find an antiferromagnetic dome accompanied by a metal-to-insulator crossover line in the intermediate-coupling regime, with a smaller superconducting dome appearing in the metallic region. Across the antiferromagnetic dome the magnetic fluctuations change from overdamped in the metallic region to propagating in the insulating region. Our findings shed new light on the intertwining between superconductivity, magnetism, and charge correlations in quantum materials.
\end{abstract}
\maketitle

\noindent \textit{Introduction.--}While the problem of interacting electrons is well-understood in the regimes where the electron-electron repulsion is much smaller or much larger than the kinetic energy, the regime where both energy scales are comparable has remained elusive. It is precisely in this regime that several unique electronic collective phenomena are observed, high-temperature superconductivity being their poster child. In the cuprates, for example, the highest superconducting (SC) transition temperatures take place as the system moves from a Mott insulating to a Fermi liquid behavior~\cite{Keimer2015}. In superconducting iron pnictides, although electronic interactions do not seem strong enough to localize the electrons, they can significantly reduce the coherence of the electronic quasi-particles~\cite{Kotliar2011,Dai2012,Georges2013}. Notwithstanding the appeal of constructing materials-specific models that can quantitatively describe and predict the properties of a moderately correlated compound, the challenges in describing this regime and its prevalence in several materials of interest warrant the investigation of minimal models that focus on key ingredients of the problem. 

The Hubbard model is perhaps the most famous such minimal model, in which electrons hopping on a lattice are subject to an onsite repulsion that mimics a strongly screened Coulomb interaction. In face of the difficulties in analyzing the intermediate-coupling regime analytically, numerical methods such as Dynamical Mean-Field Theory (DMFT)~\cite{Haule2007,Park2008,Gull2008,Weber2010}, Density Matrix Renormalization Group (DMRG)~\cite{White1992,Noack1994,Jiang2019}, or Quantum Monte Carlo (QMC)~\cite{Blankenbecler1981,Maier2005,Maier2006,Varney2009,LeBlanc2015,Ayral2015,Zheng2017,Huang2017} have been extensively applied. The main advantage of the latter is that it is an exact and unbiased method, and that is not limited to a one-dimensional geometry. However, it is intrinsically subject to the fermionic sign-problem~\cite{Loh1990,Wu2005}, which restricts the electronic occupation and temperature ranges that can be efficiently simulated. Another popular minimal model is the so-called spin-fermion model~\cite{Abanov2003}. In this case, the electron-electron interaction is substituted in lieu of a collective bosonic antiferromagnetic (AFM) order parameter that can be fine-tuned to quantum criticality. This is motivated by the fact that AFM order is often observed in moderately coupled quantum materials in proximity to unconventional superconductivity. It was recently realized that versions of the spin-fermion model with two electronic flavors (such as two bands) possess a symmetry that eliminates the sign-problem~\cite{berg2012}. This has led to a flurry of QMC studies of spin-fermion and related boson-fermion models, which revealed a nearly-universal enhancement of superconductivity at the bosonic quantum critical point (QCP)~\cite{Schattner2016,Xu2017,Gerlach2017,wang2017,Lederer2017,Berg2019,Yao2019}. However, in these models, the AFM order is introduced \emph{ad hoc} rather than being treated on an equal footing with SC and other electronic orders.

In this Letter, we construct a model free of the fermionic sign-problem in which we can treat all degrees of freedom on an equal footing. As a function of the strength of the electronic repulsion, we find an AFM dome intercepted by a metal-to-insulator crossover line at high temperatures. As temperature is lowered, this crossover line eventually becomes a first-order phase transition as it merges with the magnetic dome. Crucially, a superconducting dome only emerges near one edge of the AFM dome, providing valuable information about the nature of the pairing mechanism. We attribute this to the overdamped nature of the spin fluctuations in this region. In contrast, in the region where no SC emerges, spin fluctuations propagate ballistically and cannot mediate sufficient pairing attractions.

\begin{figure}
    \includegraphics[width=\columnwidth]{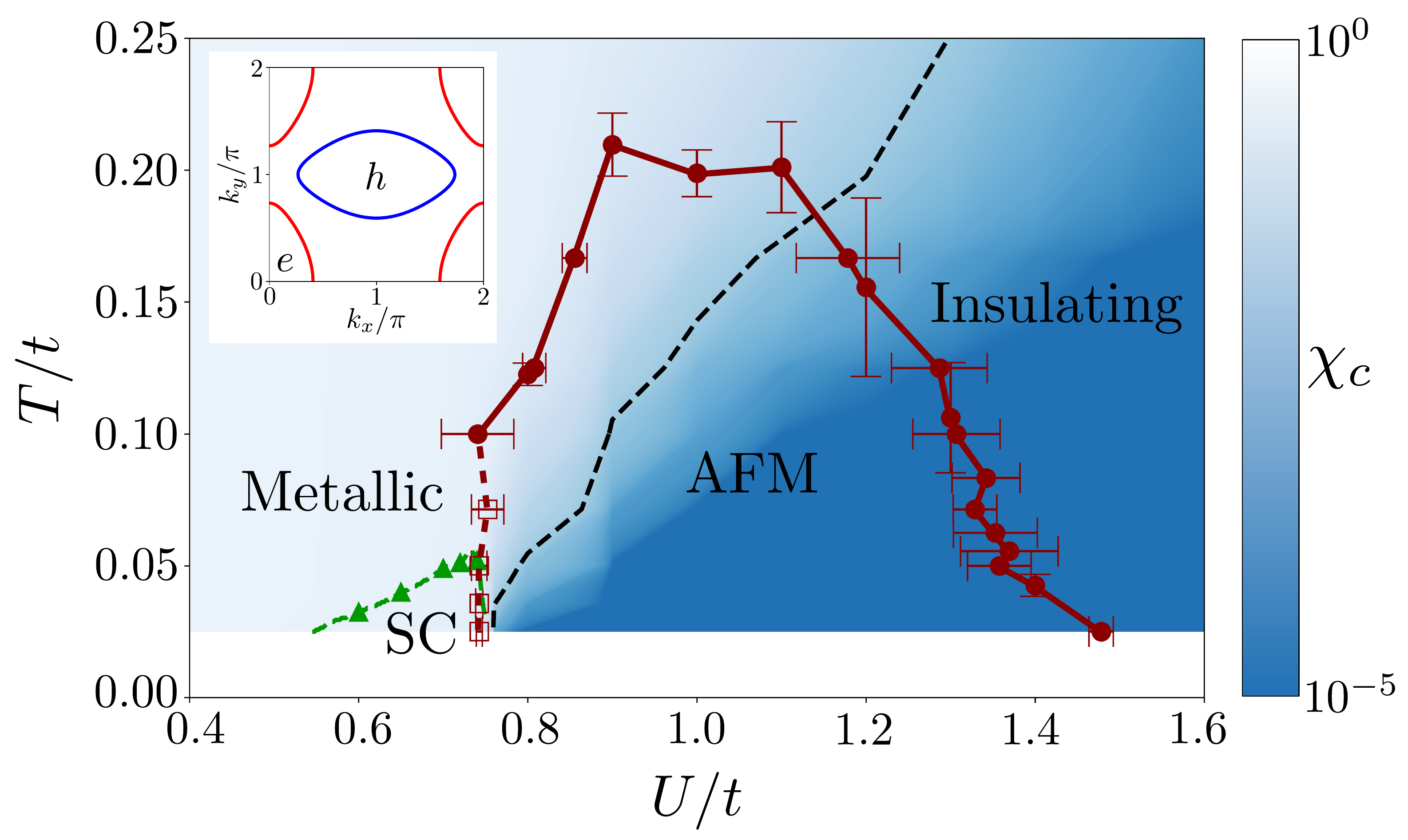}
    \caption{\textbf{Phase diagram obtained from thermodynamic observables.}
In the vicinity of $U/t\sim 1$ the phase diagram shows a variety of electronic phases, including antiferromagnetism (AFM), superconductivity (SC), and a transition between metallic and insulating behaviors. No other ordered phases were observed for $0 \leq U \leq 4t$. The dark red full circles mark the magnetic transitions determined from a scaling analysis. Near $U/t \approx 0.75$ for $T/t < 0.1$ we find that the transition becomes first order (see Supplementary), which is indicated by empty squares and a dashed red line. The color scale is logarithmic and corresponds to the compressibility, $\chi_c$ [see Fig.~\ref{fig:thermodynamics}(c)], while the black dashed line marks the contour $\chi_c=0.01$. We interpret this near complete suppression of the compressibility as a sign of insulating behavior. The green triangles mark the superconducting critical temperatures obtained from the BKT-criterion for the system size $L=12$; the green dashed line is an interpolation. The inset shows the simulated band structure, exhibiting one electron-pocket centered at $(0,0)$, and one hole pocket centered at $(\pi,\pi)$.}
\label{fig:phase_diagram}
\end{figure}

\noindent\textit{Microscopic interacting model.--}Motivated by the insight that led to the elimination of the sign-problem from spin-fermion models~\cite{Wu2005,berg2012}, we consider a simple extension of the square-lattice Hubbard model to two bands. Starting from the two-orbital Hubbard-Kanamori Hamiltonian \cite{Motome1997,Georges2013} and projecting onto states near the Fermi level, one generally obtains five distinct electron-electron interactions, $U_i$~\cite{Wu2008,Chubukov2008}. Physically, they correspond to intra-band ($U_4$, $U_5$) and inter-band ($U_1$) repulsion, spin-exchange coupling ($U_2$), and pair-exchange coupling ($U_3$). Specifically, the Hamiltonian is given by $ \mathcal{H} = \mathcal{H}_0 + \mathcal{H}_{\rm int}$, with:
\begin{align}
    \mathcal{H}_0 &= \sum_{\mathbf{k}\alpha} \epsilon^c(\mathbf{k})c^{\dagger}_{\mathbf{k}\alpha}c_{\mathbf{k}\alpha} + \sum_{\mathbf{k}\alpha} \epsilon^d(\mathbf{k})d^{\dagger}_{\mathbf{k}\alpha}d_{\mathbf{k}\alpha}  \\
    \mathcal{H}_{\rm int} &= \sum_{i\alpha\beta}\big[  U_{1} c^{\dagger}_{i\alpha}c_{i\alpha}d^{\dagger}_{i\beta}d_{i\beta} + U_{2} c^{\dagger}_{i\alpha}d^{\dagger}_{i\beta}c_{i\beta}d_{i\alpha} \nonumber \\ & + \frac{U_3}{2}\left(c^{\dagger}_{i\alpha}c^{\dagger}_{i\beta}d_{i\beta}d_{i\alpha} + \text{h.c.} \right) \nonumber \\  &+ U_{4}c^{\dagger}_{i\alpha}c^{\dagger}_{i\beta}c_{i\beta}c_{i\alpha} + U_{5}d^{\dagger}_{i\alpha}d^{\dagger}_{i\beta}d_{i\beta}d_{i\alpha} \big]\,,
\end{align}
where the operators $c$ and $d$ refer to the two bands, $\alpha$ and $\beta$ are spin indices, and $i$ and $\mathbf{k}$ are, respectively, real- and momentum-space indices. The square-lattice band dispersions $\epsilon^{c,d}(\mathbf{k})= -2(t \pm \delta)\cos k_x a - 2 (t\mp \delta) \cos k_y a \mp \mu$ are parameterized by the nearest-neighbor hopping coefficient, $t$, a hopping anisotropy $\delta$, and the chemical potential, $\mu$, see inset in Fig.~\ref{fig:phase_diagram}. Here we set $\delta=0.4t$, $\mu=-2t$, and the lattice parameter to $a=1$. As we show in the Supplementary Material, this Hamiltonian is amenable to sign-problem free QMC simulations if we consider only inter-band interactions, i.e. $U_4 = U_5 = 0$, impose the relations $U_1/4 = U_2/2 = U_3/2 = U >0$, and constrain the spin indices in the $U_1$ term to $\beta = \alpha$. This latter constraint can be interpreted as a ``single-ion'' spin anisotropy, which, in addition to allowing sign-problem free QMC simulations to be carried out, also allows for magnetic order to be stabilized at finite temperatures. Under these conditions, the Hamiltonian can be rewritten as:
\begin{equation}
    \mathcal{H} = \mathcal{H}_0 - U \sum_{i} {S_i^z} {S_i^z}\,,
\end{equation}
where
\begin{equation}
    {S_i^z} = c^{\dagger}_{i\alpha}\sigma^{z}_{\alpha\beta}d_{i\beta} + \text{h.c.}\,.
\end{equation}
Note that the above constraints are much less severe than the particle-hole symmetry that has to be imposed on the single-band Hubbard model to avoid the sign-problem. In contrast, here there are no restrictions on the electron filling of each band or on their dispersions. Importantly, as we show below, the inter-band interactions alone are sufficient to drive a plethora of ordered phases typically seen in quantum materials of interest, such as insulating behavior, magnetism, and superconductivity.

The inset in Fig.~\ref{fig:phase_diagram} depicts the specific band structure used in this work, consisting of elliptical electron- and hole-like bands at the center and at the corner of the Brillouin zone. This dispersion was chosen so that the Hamiltonian is invariant under four-fold rotations followed by particle-hole exchange and a $(\pi,\pi)$ translation in momentum space. The choice of parameters implies $\langle n^c_{i}+n^d_i \rangle=2$ but $n^c_i \neq n^d_i$, where $n^{c,d}_i$ is the electronic density of $c$ ($d$) electrons at site $i$. The elliptical shape of the Fermi surfaces was selected to suppress nesting that would otherwise favor AFM. While we performed extensive QMC simulations only for this set of band parameters, simulations over narrower parameter ranges were also performed for modified band parameters, yielding similar phase diagrams.

\begin{figure}
\includegraphics[width=\columnwidth]{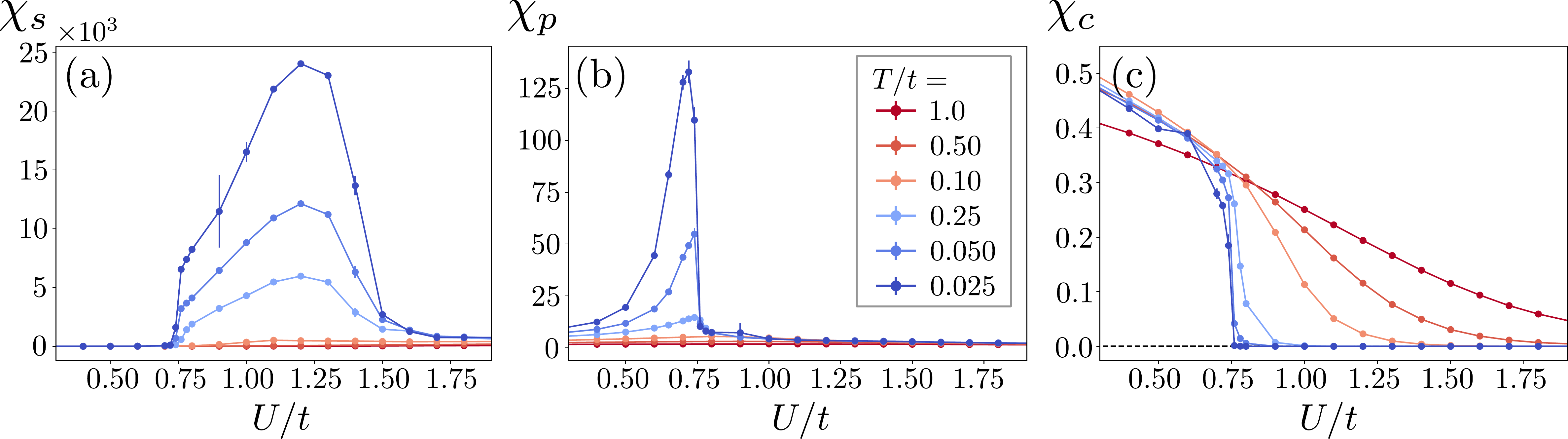}
\caption{\label{fig:thermodynamics} \textbf{Thermodynamic observables in the vicinity of $U/t \sim 1$.} (a) AFM spin-, (b) pair-, and (c) charge-susceptibilities (denoted by $\chi_s$, $\chi_p$, and $\chi_c$, respectively) for different temperatures as a function of $U/t$. The pair susceptibility peaks in the immediate vicinity of the AFM transition. At low temperatures, within our resolution, we cannot separate the transition to the AFM phase from the crossover to the insulating phase.}
\end{figure}

\noindent\textit{Phase diagram.--}The phase diagram of the microscopic model, shown in Fig.~\ref{fig:phase_diagram}, was obtained from determinant QMC simulations on $L\times L$ lattices with $L=8, 10, 12, 14$ and for temperatures $T/t \geq 0.025$. Additional details of the simulation are presented in the Supplementary Material. The salient feature of the phase diagram is an antiferromagnetic dome (red curve) in the intermediate coupling regime $U \sim t$. Indeed, as shown in Fig.~\ref{fig:thermodynamics}(a), the AFM spin susceptibility at the wave-vector $\mathbf{Q}=(\pi,\pi)$, $\chi_s = 4 U^2 \langle \int d\tau {S^z}(\mathbf{Q,\tau}) {S^z}(-\mathbf{Q,\tau=0}) \rangle + 2U$, displays a sharp enhancement at low temperatures above a critical interaction strength $U/t \approx 0.75 $, followed by a smoother suppression near $U/t \approx 1.5 $. The AFM phase boundary in Fig. \ref{fig:phase_diagram} was determined using standard finite-size scaling analysis appropriate for an Ising-type transition considering the pairs of system sizes $L=(8,12)$ and $L=(10,14)$~\cite{Toldin2015}. For $T/t<0.1$ we find evidence that the magnetic transition becomes first-order near $U/t \approx 0.75$ (see Supplementary Material). At higher temperatures and interaction strengths, the magnetic transition appears continuous. For $U/t > 1.5$, no AFM transition was observed down to the lowest temperature probed. We verified that even in the non-magnetic state, the magnetic susceptibility remains peaked at the AFM wave-vector $\mathbf{Q}=(\pi ,\pi)$.

In addition to the AFM dome, we also found a much narrower SC dome in the vicinity of $U/t=0.75$, i.e. near one of the putative AFM quantum phase transitions. The green triangles and green dashed line denote the SC transition temperatures $T_c$ as determined by the Berezinskii-Kosterlitz-Thouless (BKT) criterion, $\rho_s(T_c) = \frac{2T_c}{\pi}$ for $L=12$, interpolated between neighboring points. Importantly, this is an unconventional SC state with gaps of opposite signs in the two bands. Fig. \ref{fig:thermodynamics}(b) shows the behavior of the corresponding pair susceptibility, $\chi_p=L^{-2}\sum_{ij}\int d\tau \langle P_{\pm,i}^\dagger(\tau)P_{\pm,j}(0)\rangle$, where $P_{\pm,i} = 2(c_{i\uparrow} c_{i\downarrow} -d_{i\uparrow} d_{i\downarrow})$, as a function of $U$ and $T$. Its main features are the sharp peak observed slightly below $U/t=0.75$, where the AFM dome begins, and the absence of any enhancement near $U/t=1.5$, where the AFM dome ends. Within our resolution, the transition between the SC and AFM states appears first-order. The sharp suppression in the pair susceptibility indicates that any coexistence of the two phases is limited to a narrow range of $U/t$ in the vicinity of $U/t \approx 0.75$, although we observe no such coexistence within our resolution.

To shed light on the behavior of the charge degrees of freedom across the phase diagram, we extracted the charge compressibility, $\chi_c=L^{-2} \int d\tau \sum_{ij}\left\langle \delta\rho_i(\tau) \delta\rho_j(0)\right\rangle$, where $\delta\rho_i = n_i^c + n_i^d -2$. As shown in Fig.~\ref{fig:thermodynamics}(c), for $U/t \approx 0.75$, when AFM order sets in, $\chi_c$ displays a sudden drop at low temperatures from a finite value, indicative of a metal, to a vanishingly small value, which is indicative of an insulator. In Fig.~\ref{fig:phase_diagram} the color scale corresponds to the logarithm of $\chi_c$, clearly demonstrating a sharp transition from a metallic to an insulating phase around $U/t \approx 0.75$ at low temperatures, and a smoother crossover at higher temperatures. The black dashed line denotes the contour $\chi_c=0.01$.
The fact that the compressibility jumps sharply at low temperatures but decreases smoothly at higher temperatures supports the presence of a first-order transition between the SC phase and the AFM-insulating phase, ending in a critical endpoint followed by a Widom crossover line, as is expected for a Mott transition at finite temperatures~\cite{Terletska2011}. The precise location of the endpoint cannot be pinpointed with our available resolution.

\begin{figure}
\includegraphics[width=\columnwidth]{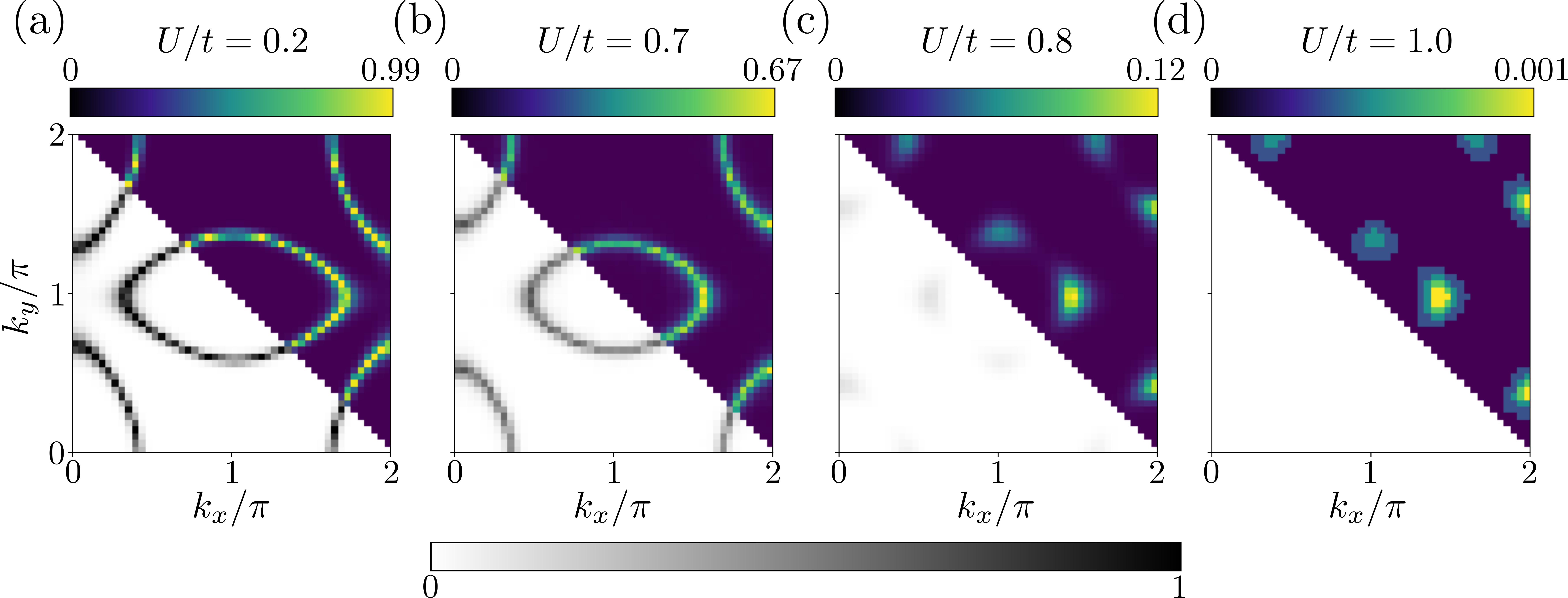}
\caption{\textbf{Evolution of the quasi-particle spectral weight proxy, $\tilde{Z}_{\mathbf{k}}$, with interaction strength.} In the upper-right-half (lower-left-half) of the panels, we plot $\tilde{Z}_{\mathbf{k}}$ for $T/t=0.05$ in a color scale (gray scale) from $0$ to $\tilde{Z}_{\rm max}$ ($0$ to $1$). Note that the two halves are identical and only the color schemes differ to highlight the loss of spectral weight across the magnetic transition. For small values of the interaction, the quasi-particle spectral weight matches the non-interacting Fermi surface shown in Fig.~\ref{fig:phase_diagram}. For larger values, the Fermi surface shrinks and, beyond $U/t=0.75$, is reconstructed, signaling the onset of AFM order with wave-vector $\mathbf{Q}=(\pi ,\pi)$. To produce these figures, we averaged over 16 different twisted boundary conditions.}
\label{fig:fermionics}
\end{figure}

\noindent\textit{Electronic and magnetic spectra.--}To further probe the impact of the metal-to-insulator crossover in the phase diagram of Fig.~\ref{fig:phase_diagram}, we extracted the electronic Green's function $\mathcal{G}$ at long imaginary time $\tau$, $\mathcal{G}_{\mathbf{k}}(\tau = \beta/2)$. Here, $\beta \equiv 1/T$ is the inverse temperature. At zero temperature and on the Fermi surface, the quantity $\tilde{Z}_{\mathbf{k}} = 2\mathcal{G}_{\mathbf{k}}(\tau = \beta/2)$ is a proxy for the quasi-particle spectral weight~\cite{Trivedi1995,Gerlach2017}, being equal to $1$ for a non-interacting system and $0$ for an insulator.  Figure~\ref{fig:fermionics} presents $\tilde{Z}_{\mathbf{k}}$ for representative values of the interaction $U$ and for a low temperature $T/t=0.05$. In each panel, the upper-right-half shows the relative spectral weight and the color scale extends to $\tilde{Z}_{\rm max}$, whereas the lower-left-half shows the absolute spectral weight and the color scale extends to $1$.

Focusing first on the upper-half of the panels, we note two effects upon increasing $U$. At $U/t = 0.7$, we see a shrinking of the Fermi surface areas, reminiscent of the so-called $s^{\pm}$-Pomeranchuk effect in multi-band systems approaching an AFM instability~\cite{Ortenzi2009,Khodas2016}. At $U/t=0.8$ and $U/t=1.0$, we observe a Fermi surface reconstruction typical of long-range AFM order, as resulting from the folding of the Brillouin zone by the AFM wave-vector $(\pi ,\pi)$. Focusing now on the lower-half of the panels, we see a strong reduction of the intensity of $\tilde{Z}_{\mathbf{k}}$ as $U$ increases, signaling a sharp suppression of the quasi-particle spectral weight. In particular, for $U/t=1.0$, the spectral weight has decreased to the point of almost vanishing, such that, for higher values of $U/t$, a Fermi surface can be barely defined. This loss of quasi-particle coherence is consistent with the suppression in the charge compressibility seen in Fig.~\ref{fig:thermodynamics}(c).

\begin{figure}
\centering
\includegraphics[width=\columnwidth]{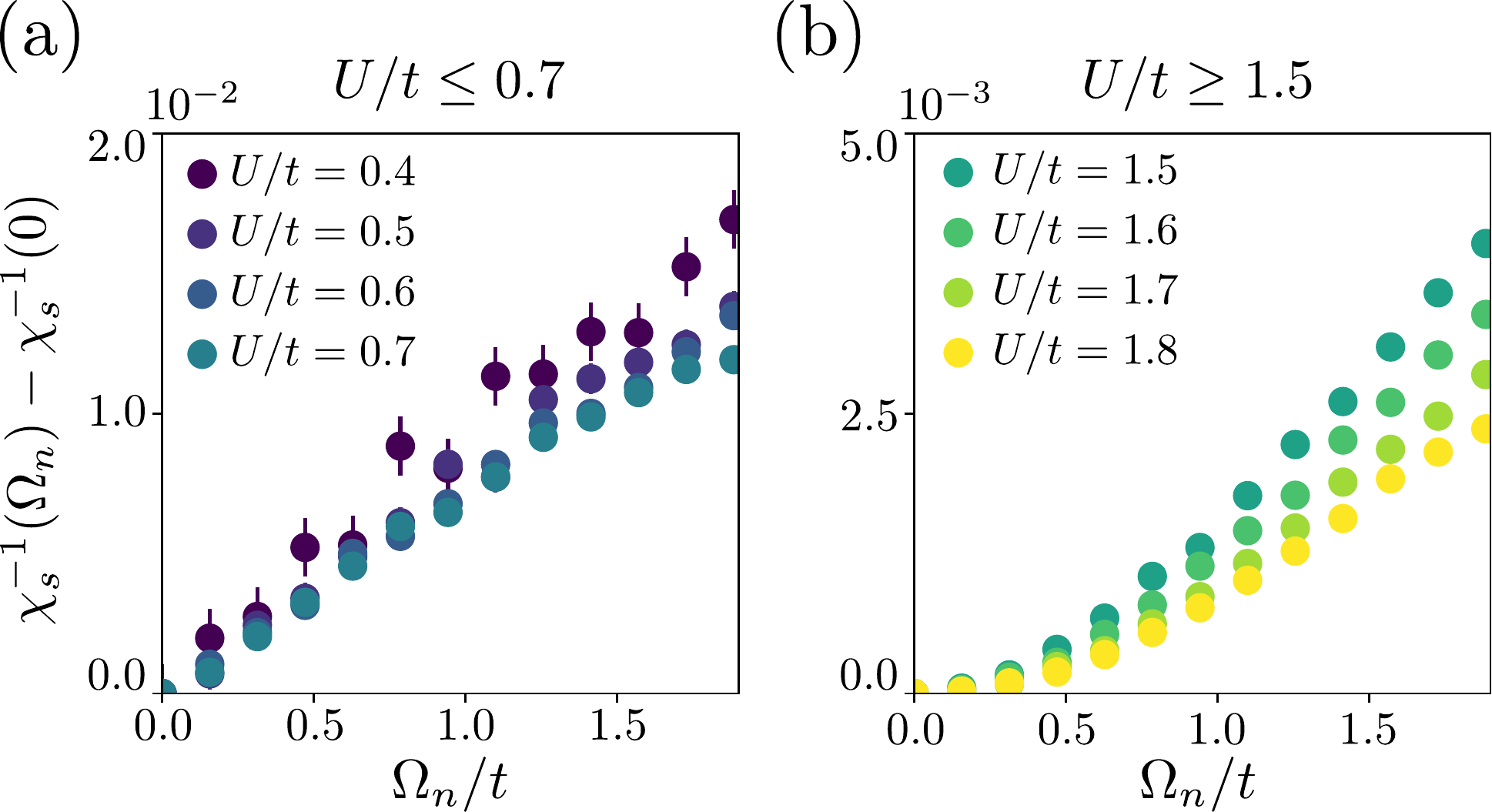}
    \caption{\textbf{Inverse dynamical spin susceptibility $\chi_s^{-1}(\Omega_n)$ in the metallic and insulating paramagnetic regions.} (a) In the regime $U/t \leq 0.7$ (i.e. to the left of the AFM dome), where the system is metallic, the dependence on $\Omega_n$ is roughly linear, $\chi^{-1}(\Omega_n) \propto \Omega_n$, indicating that the magnetic fluctuations are overdamped. (b) In the regime $U/t \geq 1.5$ (i.e. to the right of the AFM dome), where the system is insulating, the magnetic fluctuations propagate ballistically, $\chi^{-1}(\Omega_n) \propto \Omega_n^2$.}
    \label{fig:bosons}
\end{figure}

The reduction of the quasi-particle spectral weight has a drastic effect on the magnetic fluctuation spectrum in the paramagnetic state. Prior to the onset of AFM order, the electrons are reasonably coherent, as shown in Figs.~\ref{fig:fermionics}(a) and (b). The corresponding dynamic magnetic susceptibility at the AFM wave-vector, $\chi_s^{-1}(\Omega_n)$, is shown in Fig.~\ref{fig:bosons}(a) as a function of the Matsubara frequency $\Omega_n = 2 n \pi T$. In this regime, corresponding to the left of the AFM dome, the spin dynamics is overdamped, as indicated by the linear dependence $\chi_s^{-1}(\Omega_n) \sim |\Omega_n|$. This is the expected behavior arising from the decay of AFM fluctuations into collective particle-hole excitations near the Fermi surface, called Landau damping. Note that we do not expect signatures of the superconducting gap to appear here, as even the first non-zero Matsubara frequency is comparable to $T_c$. On the other hand, for $U/t \geq 1.5$, to the right of the AFM dome, the quadratic behavior $\chi_s^{-1}(\Omega_n) \sim \Omega_n^2$ shown in Fig.~\ref{fig:bosons}(b) is typical of ballistic spin dynamics, with AFM fluctuations propagating without damping. The fact that the quasi-particle spectral weight is strongly reduced for $U/t > 0.75$ suggests that this absence of damping is a consequence of the suppression of the decay channel of an AFM excitation into quasi-particles.

\noindent\textit{Discussion.--}Having completely characterized the phase diagram of the electronic two-band model shown in Fig.~\ref{fig:phase_diagram}, we now discuss its implications for our understanding of the intertwining between AFM and SC in the intermediate coupling regime. The appearance of an AFM dome can be rationalized by interpolating the expected behaviors in the metallic and insulating sides of the phase diagram. From a weak-coupling perspective, because the electron-like and hole-like bands are not nested, the interaction strength must overcome a threshold value for AFM order to onset. From a strong-coupling perspective, the two-band model maps onto an Ising model with a strong transverse field (see Supplementary Material), and as a result the ground state is a featureless, insulating quantum paramagnet. What is surprising, however, is the fact that the threshold value for $U/t$ at the lowest temperature probed coincides (within our resolution) with the value that triggers a metal-to-insulator transition, characterized by vanishing compressibility and quasi-particle spectral weight.
Additionally, we note that the onset of AFM order is not due to Fermi surface nesting, as  numerical simulations of two electron-like band dispersions (not shown) also reveal a magnetic dome at similar values of $U/t$.

The numerical results suggest the presence of two putative AFM quantum phase transitions near $U/t \approx 0.75$ and $U/t \approx 1.5$. Of course, the AFM transition temperature could remain non-zero beyond this range, since the lowest temperature that we probe is $T/t = 0.025$. Although this makes it difficult to locate a possible QCP, the fact that the AFM susceptibility is strongly suppressed for these two values of the interaction strength [as shown in Fig.~\ref{fig:thermodynamics}(a)] allows us to make a meaningful comparison between them. The main difference is that long-range superconductivity appears near $U/t=0.75$, while not even weak SC fluctuations are observed near $U/t=1.5$. Thus, while this result supports the point of view that AFM fluctuations play an important role in promoting high-temperature superconductivity -- the highest $T_c$ in our system is a few percent of $t$ -- it also makes it clear that proximity to an AFM transition is by no means enough for superconductivity to be triggered. On the contrary, our analysis of the spin dynamics in Fig.~\ref{fig:bosons} reveals that overdamped (i.e. ``slow") fluctuations are much better at promoting Cooper pairing than ballistic (i.e. ``fast") fluctuations. This change in the character of the spin dynamics, in turn, can be attributed to the strong suppression of the quasi-particle spectral weight shown in Fig.~\ref{fig:fermionics}, which effectively eliminates Landau damping. It is important to note that, despite the quasi-particle spectral weight being heavily suppressed, as long as it remains finite at non-zero temperatures, superconductivity could in principle still arise~\cite{wang2016}.\newline

\noindent\textit{Conclusion.--}In conclusion, we demonstrated that a suitable two-band version of the Hubbard model can be efficiently simulated via QMC without the fermionic sign-problem. The resulting phase diagram showcases various ordered states typically found in quantum materials, such as AFM, SC, and a correlated insulating phase. More importantly, our results offer an unbiased view of the rich interplay between these different degrees of freedom, demonstrating that both AFM and SC are enhanced near the metal-to-insulator transition in the intermediate-coupling regime. Future investigations of this type of model would be desirable to shed light on the fermionic properties near the onset of the AFM order, particularly to elucidate whether non-Fermi liquid behavior or pseudogap behavior are also triggered by inter-band repulsive interactions.

\begin{acknowledgments}
We thank A. Chubukov, A. Klein, Z. Y. Meng, and O. Vafek for fruitful discussions. MHC and RMF are supported by the U.S. Department of Energy, Office of Science, Basic Energy Sciences, Materials Science and Engineering Division, under Award No. DE-SC0020045. RMF also acknowledges partial support from the Research Corporation for Science Advancement via the Cottrell Scholar Award. XW acknowledges financial support from National MagLab, which is funded by the National Science Foundation (DMR-1644779) and the state of Florida. YS was supported by the Department of Energy, Office of Basic Energy Sciences, under contract no. DE-AC02-76SF00515 at Stanford, by the Gordon and Betty Moore Foundation's EPiQS Initiative through Grant GBMF4302 and GBMF8686 and by the Zuckerman STEM Leadership Program. EB was supported by the European Research Council (ERC) under grant HQMAT (grant no. 817799), the US-Israel Binational Science Foundation (BSF), the Minerva foundation, and a research grant from Irving and Cherna Moskowitz. We thank the Minnesota Supercomputing Institute (MSI) at the University of Minnesota, where a part of the numerical computations was performed.
\end{acknowledgments}

\bibliographystyle{apsrev4-1}

\bibliography{scibib_revtex}

\clearpage

\begin{widetext}

\begin{center}
\textbf{Supplementary material for ``Modeling unconventional superconductivity at the crossover between strong and weak electronic interactions"}
\end{center}

\setcounter{equation}{0}
\renewcommand{\theequation}{S\arabic{equation}}

\setcounter{figure}{0}
\renewcommand{\thefigure}{S\arabic{figure}}

\setcounter{section}{0}
\renewcommand{\thesection}{S\Roman{section}}

\section{Model and Determinant Quantum Monte Carlo}

In this work, we perform determinant Quantum Monte Carlo (DQMC) simulations of the following model:
\begin{equation}
    H = \sum_{\mathbf{k}\alpha}\left(\epsilon^c(\mathbf{k}) c^{\dagger}_{\mathbf{k}\alpha}c_{\mathbf{k}\alpha} + \epsilon^d(\mathbf{k}) d^{\dagger}_{\mathbf{k}\alpha}d_{\mathbf{k}\alpha}\right) -U \sum_{i}S^{z}_{i}S^{z}_{i}\,,
\end{equation}
where $S^{z}_i = c^{\dagger}_{i\alpha}\sigma^{z}_{\alpha\beta}d_{i\beta} + \text{h.c.}$. The dispersions are
\begin{equation}
	\epsilon^{c/d}(\mathbf{k}) = -2(t \pm \delta)\cos k_x a - 2 (t\mp \delta) \cos k_y a \mp \mu\,,
\end{equation}
and we choose $\delta=0.4t$, $\mu=-2t$, and $a=1$. Expanding the four-fermion term $U \sum_{i}S^{z}_{i}S^{z}_{i}$ and using the standard fermionic anti-commutation relations, we find
\begin{equation}
    \mathcal{H}_{\rm int} = U \sum_{i\alpha\beta} \left[4 c^{\dagger}_{i\alpha}c_{i\alpha}d^{\dagger}_{i\beta}d_{i\beta}\delta_{\alpha\beta} + 2c^{\dagger}_{i\alpha}d^{\dagger}_{i\beta}c_{i\beta}d_{i\alpha} + \left( c^{\dagger}_{i\alpha}c^{\dagger}_{i\beta}d_{i\beta}d_{i\alpha} + \text{h.c.} \right) \right]\,,
\end{equation}
Thus, we obtain the same expression as Eq. (2) of the main text with vanishing intra-band interactions ($U_4 = U_5 = 0$), inter-band interactions $U_1 = 4U$, $U_2 = 2U$ and $U_3 = 2U$, and $\alpha=\beta$ in the first term.

In the DQMC technique, after discretizing imaginary time into $N_\tau$ slices, the interaction term is decoupled by introducing a Hubbard-Stratonovich field. The partition function is then evaluated by statistical sampling of the field configurations~\cite{Blankenbecler1981}, with a weight given by the determinant of the fermionic Green's function (see Eq.\eqref{eq:partition_function}). In the present case, we apply the discrete Hubbard-Stratonovic transformation (HST) \cite{Assaad2002} to the interaction term
\begin{equation}
e^{\Delta \tau U{S_i^z}{S_i^z}}=\frac{1}{4}\sum_{l=\pm1,\pm2}\gamma(l)e^{\Delta\tau \phi(l) {S_i^z}}+O(\Delta \tau U)^{4},
\end{equation}
where $\phi(l) = \sqrt{\frac{U}{\Delta\tau}}\eta(l)$,
$\gamma(\pm1)=1+\sqrt{6}/3$, $\gamma(\pm2)=1-\sqrt{6}/3$, $\eta(\pm1)=\pm\sqrt{2(3-\sqrt{6})}$, $\eta(\pm2)=\pm\sqrt{2(3+\sqrt{6})}$ and $\Delta \tau =\beta / N_\tau$. For technical reasons, for $0.7<U<0.8$ and $L=12$ we used a continuous HST,
\begin{equation}
e^{\Delta \tau U{S_i^z}{S_i^z}} = C \int d\phi e^{-\Delta \tau \phi^2/4U+ \Delta \tau \phi {S_i^z}},
\end{equation}
where $C$ is a constant.
The discrete HST procedure leads to a shorter autocorrelation time than the continuous HST, but has no impact on the physics.

In either decoupling scheme, the partition function can be written as
\begin{equation}
    \mathcal{Z} = \int \mathcal{D}[\phi]\text{det}\left[\widehat{G}(\phi)^{-1} \right]\exp\left(-\mathcal{S}_{\phi} \right)\,,\label{eq:partition_function}
\end{equation}
where $\text{det}\left[ \widehat{G}(\phi)^{-1} \right]$ is the inverse fermionic determinant dependent on $\phi$ and $\mathcal{S}_{\phi}$ is the $\phi$-dependent part of the action. For a given Hubbard-Stratonovich field configuration, the matrix $\hat{G}(\phi)$ commutes with an anti-unitary operator: $A = is^y\otimes \sigma^zK$ and $A^2=-1$, where $\sigma$ and $s$ are Pauli matrices acting on the band and spin subspaces respectively, and $K$ denotes complex conjugation. As discussed in an earlier work \cite{berg2012}, such an anti-unitary symmetry guarantees that the fermionic determinant is positive definite for arbitrary energy dispersions and field configurations. As a result, the DQMC algorithm does not suffer from the notorious fermion sign problem.

To minimize finite-size effects, a single quantum of a pseudo-magnetic field was inserted such that $\Phi_{c\uparrow}= \Phi_{d\downarrow} = -\Phi_{c\downarrow} = -\Phi_{d\uparrow}$, where $\Phi_{\alpha\sigma}$ is the flux felt by the fermions of band $\alpha$ and spin $\sigma$  ~\cite{Assaad2002}. We used grids of size $L \times L$ in real-space (with $L=8,10,12,14$\footnote{\parbox{\textwidth}{Only the magnetic susceptibility was measured for $L=14$, and used in the crossing analysis for determination of the magnetic phase boundaries.}}), and size $N_{\tau}$ along the imaginary time direction. The value of $N_{\tau}$ depends on temperature $T=1/\beta$) and is chosen so that the time discretization $\Delta \tau \equiv \beta/N_\tau = 0.05$. The configurations of the HST fields are generated following the Metropolis algorithm of local field updates \cite{Blankenbecler1981}. For every choice of parameters, we run 8 parallel Markov chains of 12,000 total system sweeps. The first 2,000 configurations are dropped to ensure thermal equilibration.

To estimate the statistical errors arising due to the finite thermal ensemble averaging, we first compute the auto-correlation time, $\tau_\mathcal{O}$, for each quantity $\mathcal{O}$. This is necessary since the configurations constructed using the Metropolis algorithm are not independent.
Generally, higher moments exhibit longer auto-correlation times, implying that the auto-correlation time of e.g. the spin-spin correlation time is longer than the auto-correlation time of $\phi$. Combining the auto-correlation time with the variance of the correlated configurations ($\sigma_\mathcal{O}^2$) yields an estimate for the statistical error $\delta_\mathcal{O}$~\cite{Gubernatis2016}:
\begin{equation}
    \delta^2_\mathcal{O} = (1 + 2\tau_\mathcal{O}) \frac{\sigma_\mathcal{O}^2}{M},
\end{equation}
where $M$ is the total number of statistical configurations. For quantities where an auto-correlation time cannot be defined we instead use a jackknife procedure~\cite{Gubernatis2016}. This is the case for e.g. the compressibility which cannot be defined without reference to all configurations within the thermal ensemble.

\begin{figure}
\includegraphics[width=0.32\textwidth]{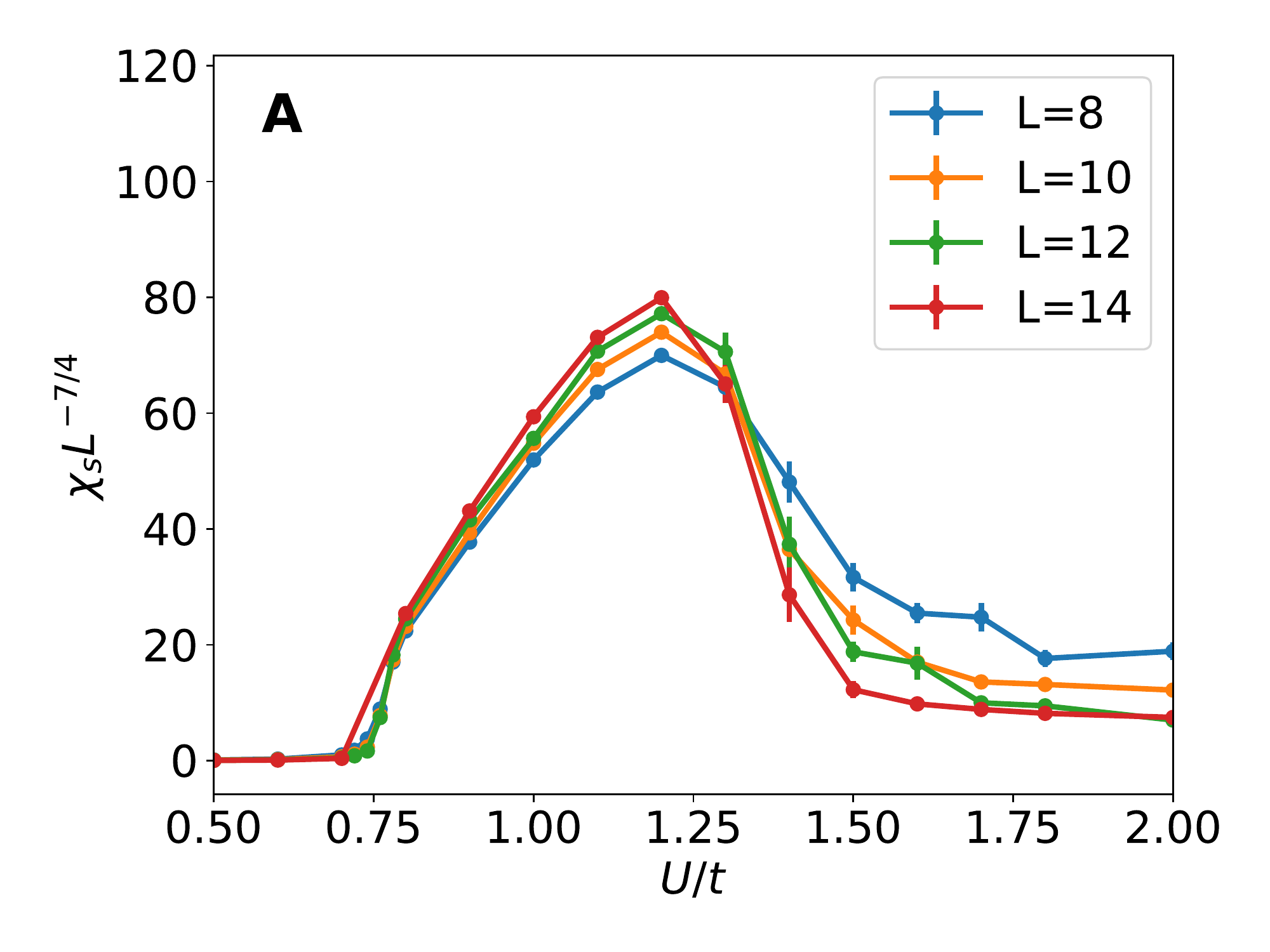}\includegraphics[width=0.32\textwidth]{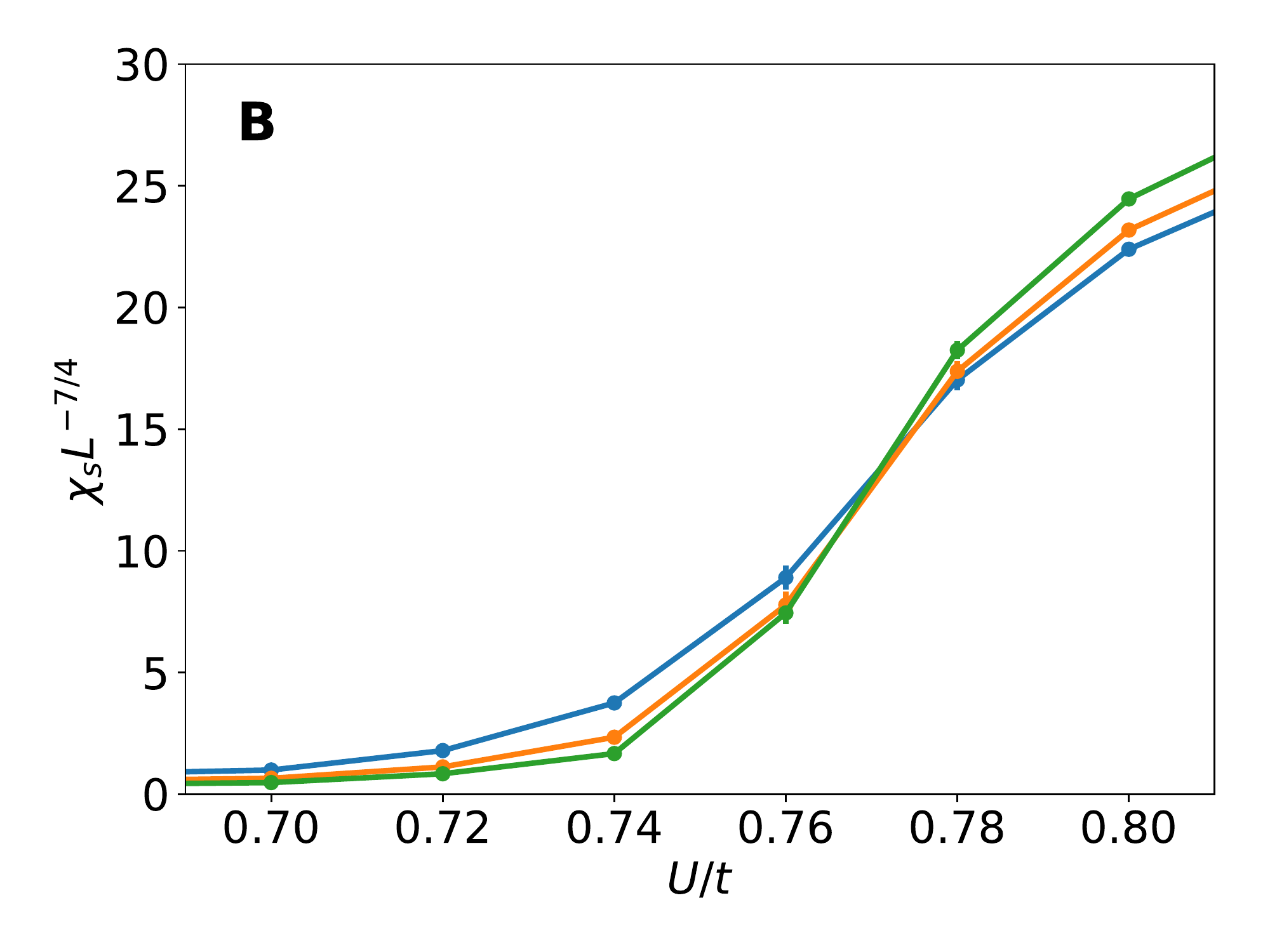}\includegraphics[width=0.32\textwidth]{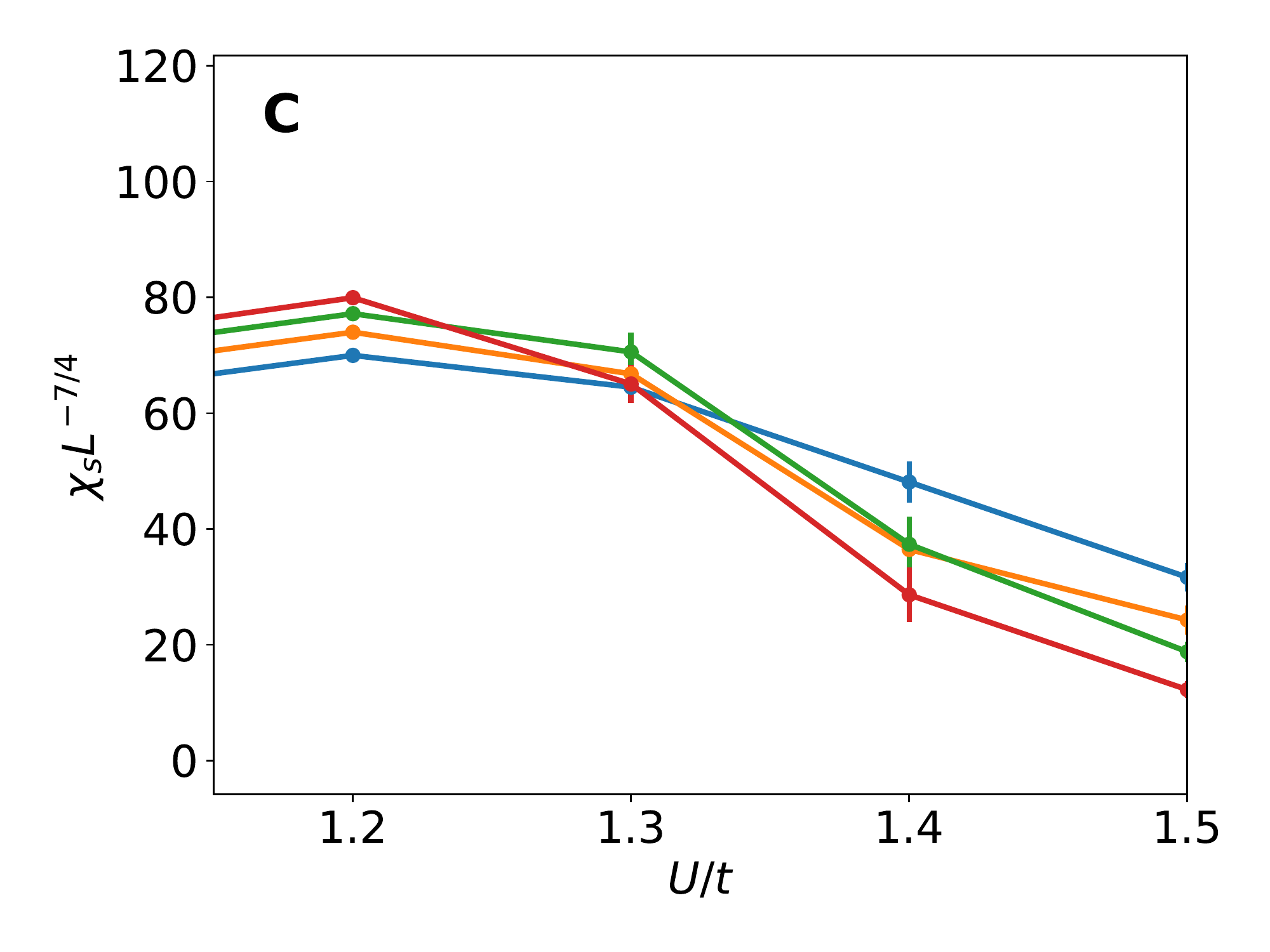}
\caption{\label{fig:spin_susc_scaling_beta_10} Scaling of the spin susceptibility $\chi_s$ with $L^{-7/4}$, for $\beta t =10$, corresponding to the critical behavior of the two-dimensional Ising model. {\sffamily{\bfseries B}} and {\sffamily{\bfseries C}} represent the zoomed-in plots near the two sides of the magnetic dome. In {\sffamily{\bfseries B}} we omitted the $L=14$ line as this was only evaluated on a coarser $U$ grid.}
\end{figure}

\begin{figure}
\includegraphics[width=0.32\textwidth]{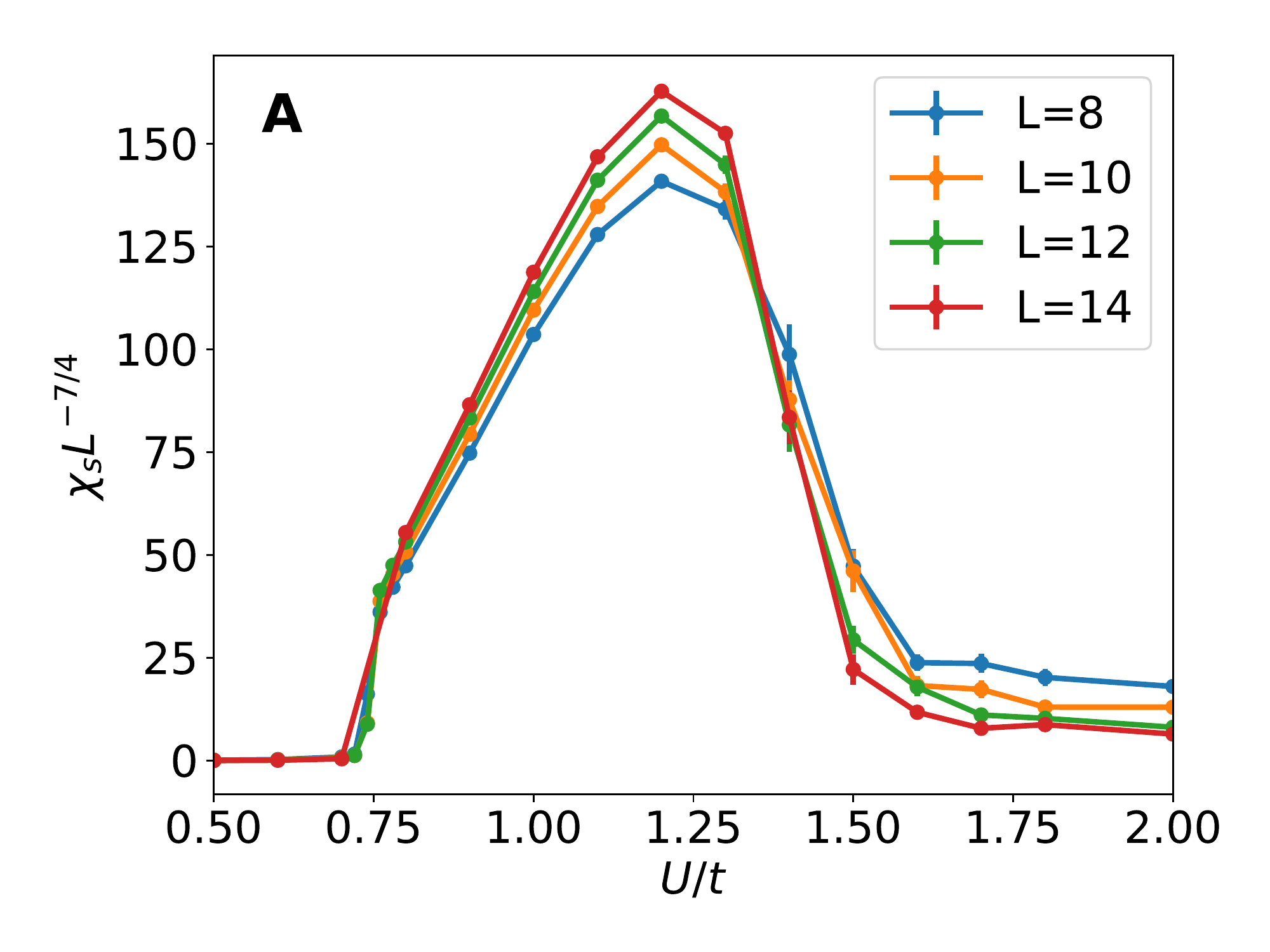}\includegraphics[width=0.32\textwidth]{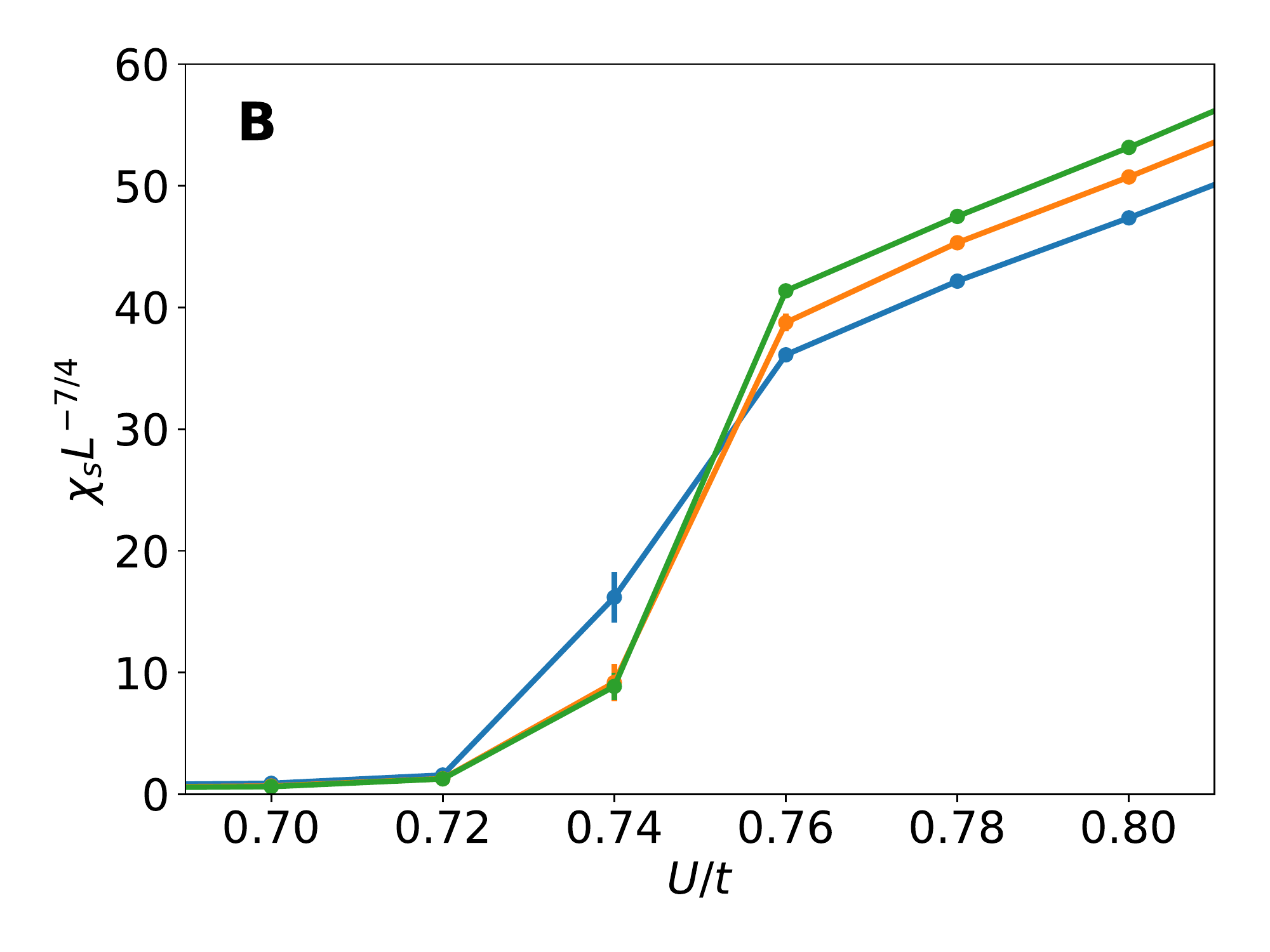}\includegraphics[width=0.32\textwidth]{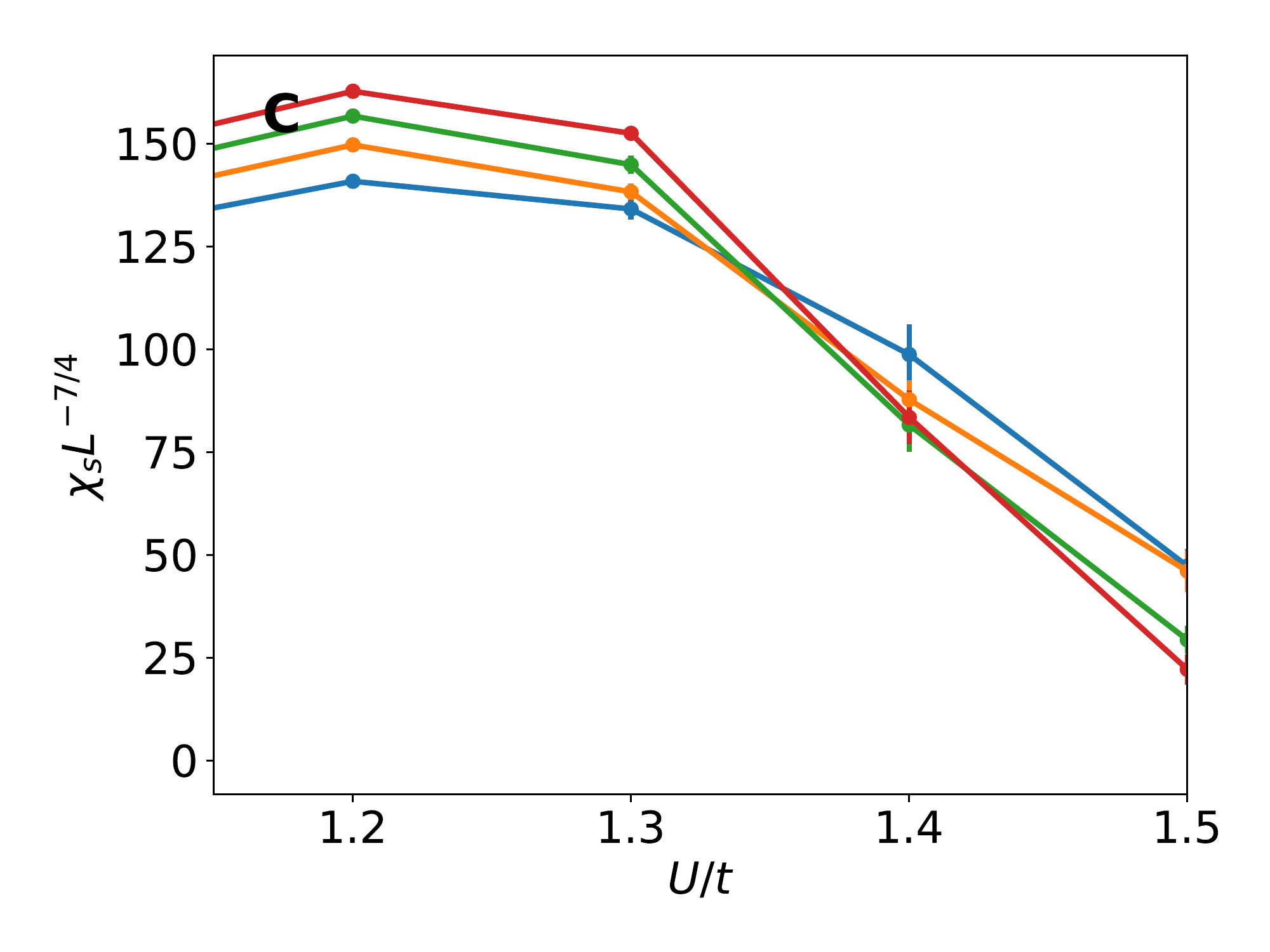}
\caption{\label{fig:spin_susc_scaling_beta_20} Scaling of the spin susceptibility $\chi_s$ with $L^{-7/4}$, for $\beta t =20$, corresponding to the critical behavior of the two-dimensional Ising model. {\sffamily{\bfseries B}} and {\sffamily{\bfseries C}} represent the zoomed-in plots near the two sides of the magnetic dome. In {\sffamily{\bfseries B}} we omitted the $L=14$ line as this was only evaluated on a coarser $U$ grid.}
\end{figure}

\section{Identifying the antiferromagnetic transition}
We locate the antiferromagnetic transition by studying the spin-spin correlation function, which is calculated from
\begin{equation}
    \chi_s(\mathbf{r}_i,\tau) = \frac{1}{L^2N_\tau}\sum_{\tau',\mathbf{r}_j}\langle \phi(\mathbf{r}_i+\mathbf{r}_j,\tau+\tau') \phi(\mathbf{r}_i,\tau') \rangle\,,
\end{equation}
where both $\mathbf{r}_i$ and $\tau$ refer to discretized variables, and $\langle \cdots \rangle$ denotes ensemble averaging. The thermodynamic spin susceptibility is defined via $\chi_s = \frac{\beta}{N_\tau}\sum_{\mathbf{r}_i,\tau} \chi_s(\mathbf{r}_i,\tau)$. Alternatively, the spin susceptibility can also be defined via the fermionic operators: $\tilde{\chi}_s =\frac{\beta}{N_\tau} \sum_{\mathbf{r}_i,\tau} \langle S^z(\mathbf{r}_i,\tau)S^z(0,0) \rangle$. There is a relation between the fermionic and bosonic susceptibilities, given by:
\begin{equation}
    \chi_s = (2U)^2\tilde{\chi}_s + 2U,
\end{equation}
which is exact in the case of the continuous HST, and is correct up to $O(\Delta\tau U)^5$ in the case of the discrete HST. This relation has been verified by our numerical results. 

Evaluating $\chi_s$ for different system sizes ranging from $L=8,\ldots,14$ allows us to carry out a finite size scaling analysis to determine the location of the magnetic phase transition. The susceptibility follows the scaling function $\chi_s(t,L)=L^{\gamma/\nu}g(t L^{1/\nu})$ \cite{fisher72}, where $t$ is the reduced temperature, and $\{\gamma,\nu\}$ are critical exponents in the thermodynamic limit. Here we limit ourselves to a simple crossing analysis, where we use the fact that, at the transition, $L^{-\gamma/\nu}\chi_s(0,L)=g(0)$. In other words, a magnetic transition occurs at points where the quantities $L^{-\gamma/\nu}\chi_s(0,L)$ computed for different values of $L$ cross, as seen in Figs.~\ref{fig:spin_susc_scaling_beta_10} and
~\ref{fig:spin_susc_scaling_beta_20}. Here, we use the standard Ising exponents for two-dimensional systems, $\gamma=7/4$ and $\nu=1$. To reduce statistical noise, we consider only the crossings between pairs of system sizes separated by a fixed $\delta L=4$~\cite{Toldin2015}, namely crossings between the $L=8$ and $L=12$ data and between the $L=10$ and $L=14$ data. This does not significantly impact the location of the crossings themselves, as seen in Fig.~\ref{fig:phase_diag_all_points}.

To estimate the error associated with the determination of the transition, we linearly interpolate between the two points on either side of the transition for two given system sizes and use that the $x$-coordinate of their intersection is given by

\begin{equation}
    x_c = x_1 + \left[ \chi_S^{L_2}(x_1)-\chi_S^{L_1}(x_1)\right] \left[\frac{\chi_S^{L_1}(x_2)-\chi_S^{L_1}(x_1)}{x_2-x_1}-\frac{\chi_S^{L_2}(x_2)-\chi_S^{L_2}(x_1)}{x_2-x_1}\right]^{-1}\,.
\end{equation}
\begin{figure}
\includegraphics[width=0.5\textwidth]{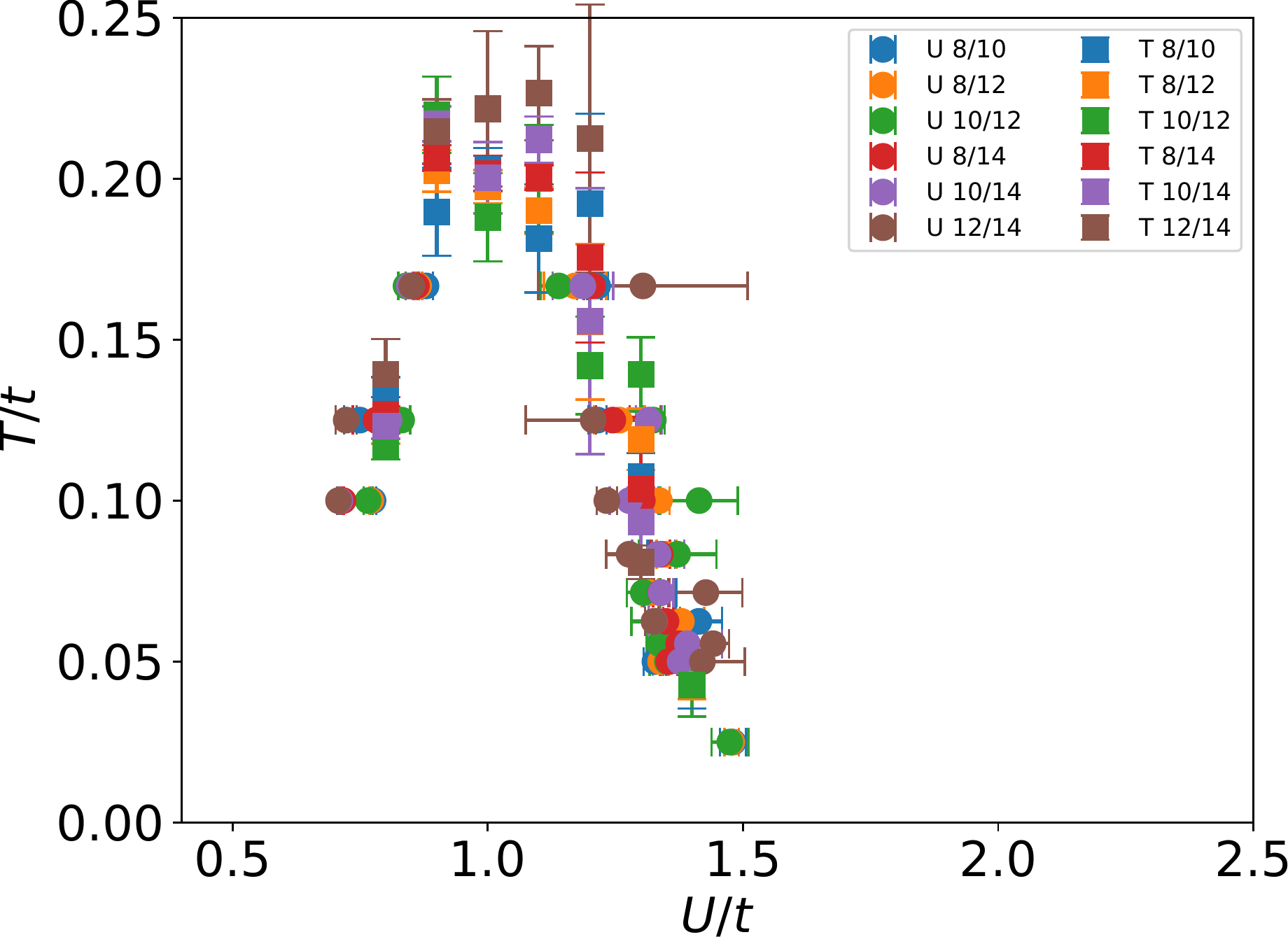}
\caption{\label{fig:phase_diag_all_points} Magnetic phase boundary based on crossing analysis of the scaled spin susceptibility $\chi_s(L)L^{-\gamma/\nu}$ as a function of both $U$ (circles) and $T$ (squares). We included only the pairs $L=(8,12)$ and $L=(10,14)$ in the crossing analysis leading to the phase diagram shown in the main text.}
\end{figure}

\begin{figure}
\includegraphics[width=0.4\textwidth]{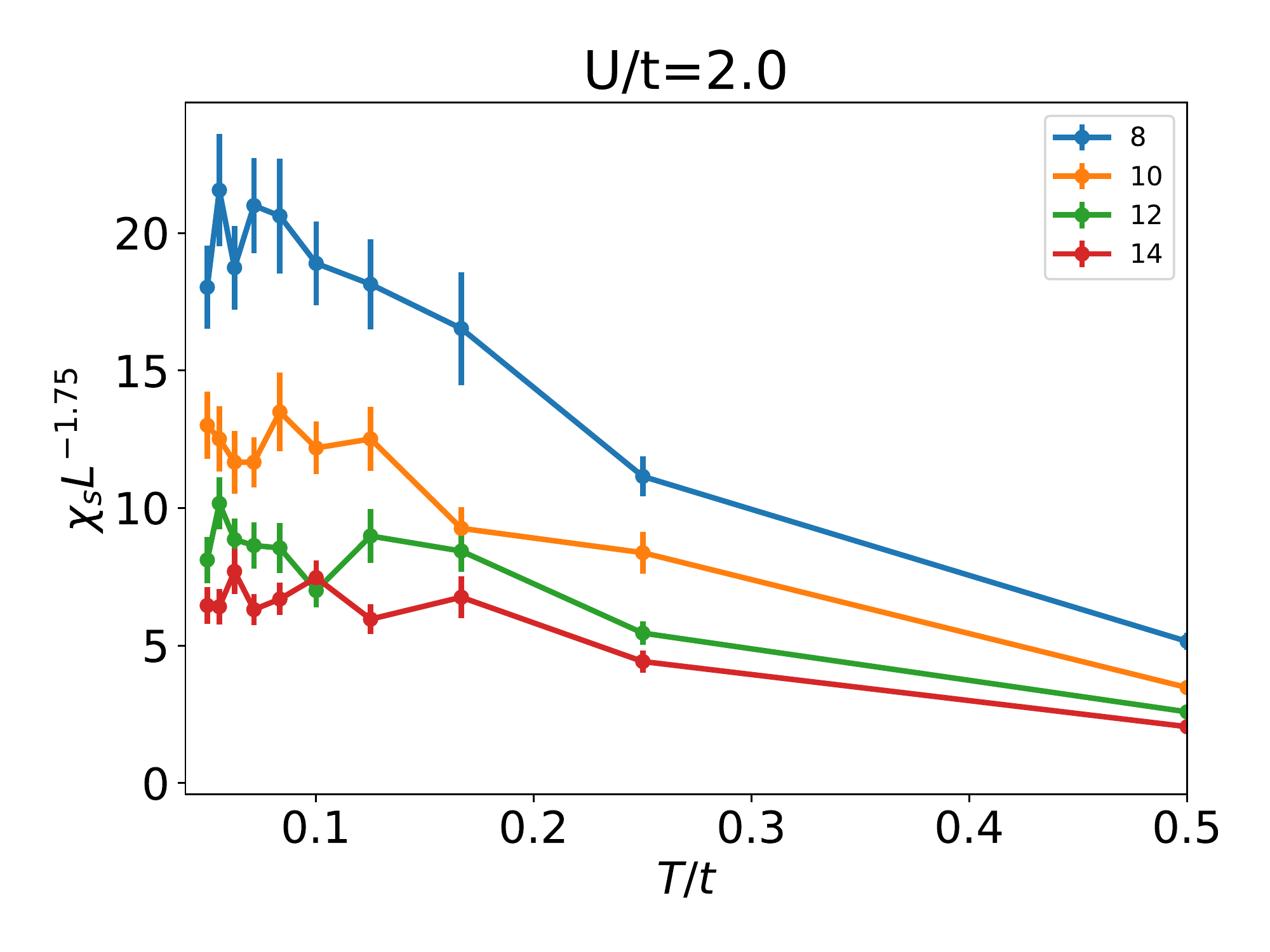}
\includegraphics[width=0.4\textwidth]{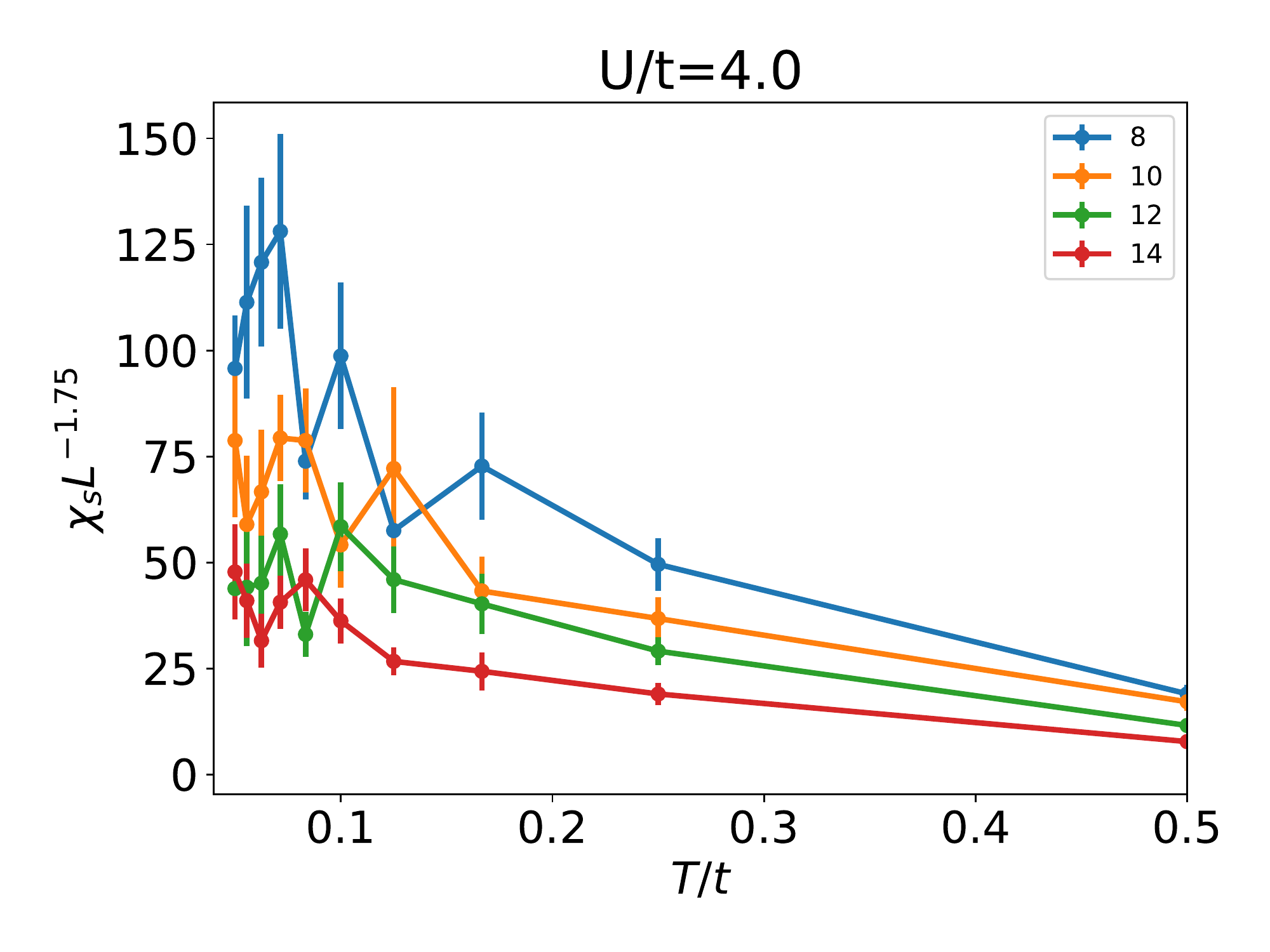}
\caption{Scaled spin susceptibilities $\chi_s L^{-7/4}$ for $U/t=2.0$ and $U/t=4.0$.  While the scaled susceptibilities do exhibit crossings, we observe that these are due to fluctuations in the data rather than actual phase transitions. Hence, these points were omitted in the above figure and in further considerations.}
\label{fig:spurious_crossings}
\end{figure}

\begin{figure}
    \centering
    \includegraphics[width=\linewidth]{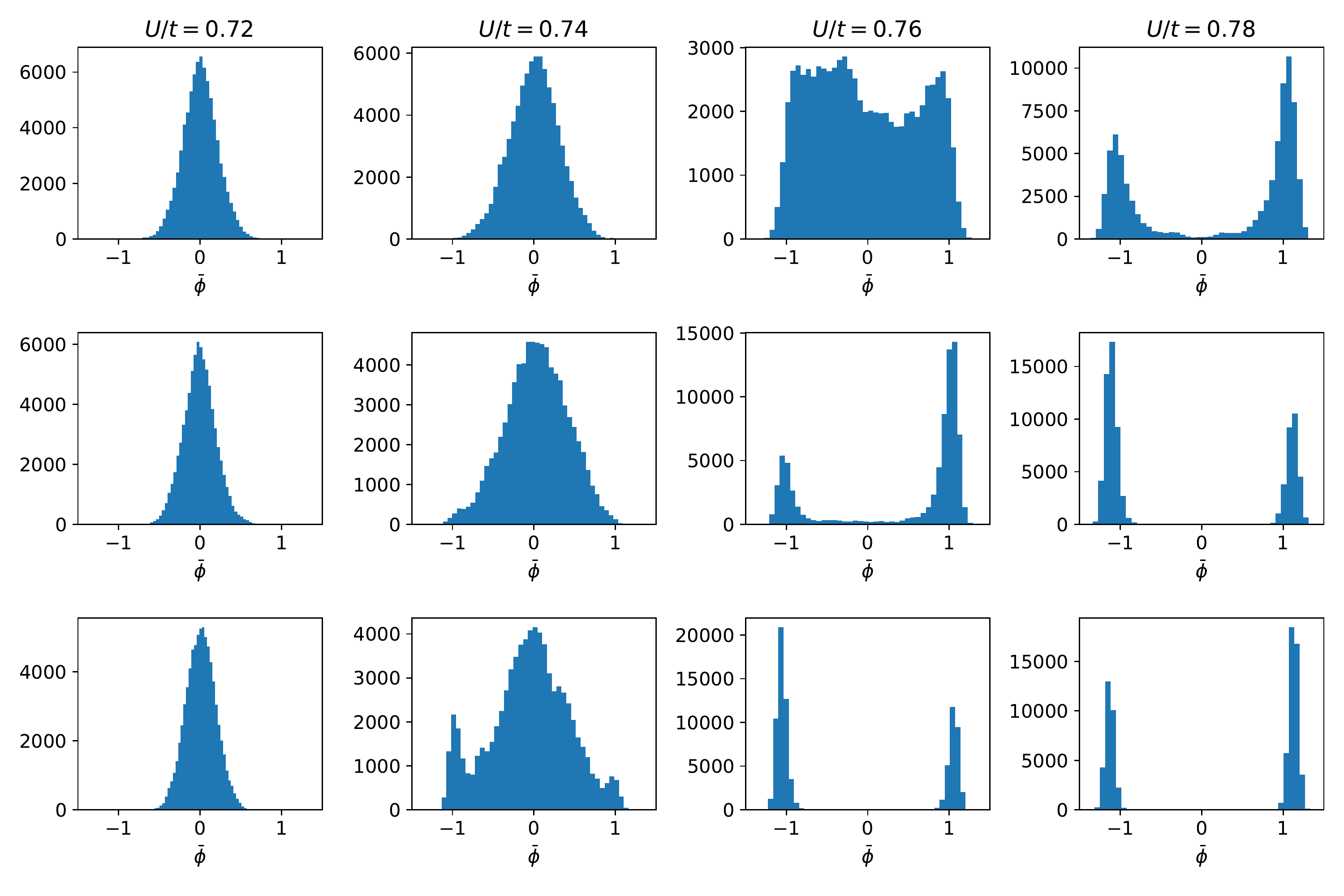}
    \caption{Histograms of the average staggered magnetization $\bar{\phi}\equiv \frac{1}{N_\tau L^2}\sum_{\mbf{r}_i,\tau}\phi(\mbf{r}_i,\tau)$ for a few values of $U/t$ and for $\beta t =10$ (upper panel), $\beta t=14$ (middle panel), and $\beta t=20$ (lower panel). Results are obtained for $L=12$. The triple-peak structure near $U/t=0.74$ at $\beta t=20$ indicates a first-order magnetic phase transition.}
    \label{fig:firstorderSDW}
\end{figure}

Standard error propagation provides an estimate for the error associated with the value of $x_c$. The results from different system sizes are averaged, and the error on this result is estimated through the quadratic sum of the variance of the estimates and the errors on the individual estimates. This procedure is carried out for crossings both along the $T$ and $U$ axes. This provides the antiferromagnetic phase boundary shown in Fig.~1{\sffamily{\bfseries C}} of the main text. Points with horizontal error bars are obtained from crossings as a function of $U$, while points with vertical error bars are obtained from crossings as a function of $T$. For completeness, in Fig.~\ref{fig:phase_diag_all_points} we show the antiferromagnetic phase transitions without averaging over different system sizes. Note that we do not have $L=14$ data below $T/t=0.05$, and thus the crossings shown at temperatures below this value are obtained from the crossing between the $L=8$ and $L=12$ data only. This analysis does not include points for which we have indications that the transition is first order, as discussed below. Note that, for large values of $U$, there are a number of spurious crossings due to fluctuations in the data, as shown in Fig.~\ref{fig:spurious_crossings}. Of course, these spurious points are not included in the phase diagram.

\begin{figure}
\centering
\includegraphics[width=1\textwidth]{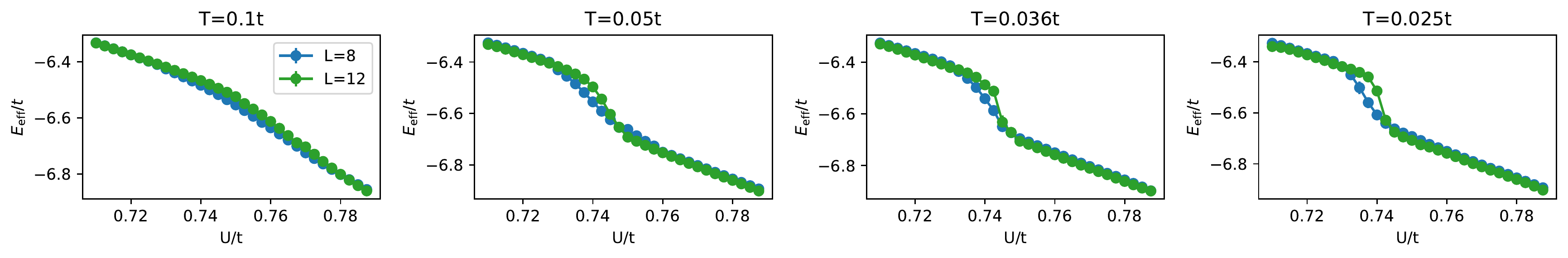}
\caption{The effective energy $E_\mathrm{eff}$ as defined in Eq.~\eqref{eq:E_eff} for different temperatures. As the system size increases at low temperatures, a sharp feature appears at the onset of magnetic order, indicating a first-order transition.
\label{fig:logdet}}
\end{figure}

To check for the possibility of a first-order transition, we begin by analyzing the histograms of the uniform staggered magnetization, $\bar{\phi}\equiv \frac{1}{L^2N_\tau}\sum_{\mbf{r}_i,\tau}\phi(\mbf{r}_i,\tau)$, shown in Fig.~\ref{fig:firstorderSDW}. At low temperatures ($\beta t = 20$) and for $U/t = 0.74$, the histogram shows a triple-peak structure, hinting at a first-order magnetic transition~\cite{Binder1984}.
To further investigate the possibility of a first-order transition, we examine the $U$ dependence of the effective energy:
\begin{equation}
E_\mathrm{eff} = \left\langle T \log(\det(\widehat G(\phi)))/L^2\right\rangle.
\label{eq:E_eff}
\end{equation}
$E_\mathrm{eff}$ is analogous to one of the energy terms in a classical Monte Carlo simulation, and is expected to show a discontinuous jump at a first order transition at $U_c$ in the thermodynamic limit. For a finite size system, the discontinuity is replaced by a sharp feature over a scale $\Delta U/U_c \approx L^{-2}$, whereas in a continuous transition no such jump should occur. As shown in Fig.~\ref{fig:logdet}, while such a feature clearly appears at low temperatures, it is not detectable within our resolution at $T/t=0.1$. For enhanced resolution in $U$, here we simulated $U/t=0.72, 0.73 ... 0.78$, and reweighted the data to obtain the rest of the values of $U$~\cite{Ferrenberg1988}. For each system size, we define $U_c(L)$ as the position of the maximum of the derivative $\frac{\partial E_\mathrm{eff}}{\partial U}$. In the phase diagram in the main text, we estimate $U_c\approx U_c(L=12)$, and the errors are obtained from the widths of Lorentzians fitted to the numerical derivatives of the data shown in Fig.~\ref{fig:logdet}.

\begin{figure}
\includegraphics[width=0.65\textwidth]{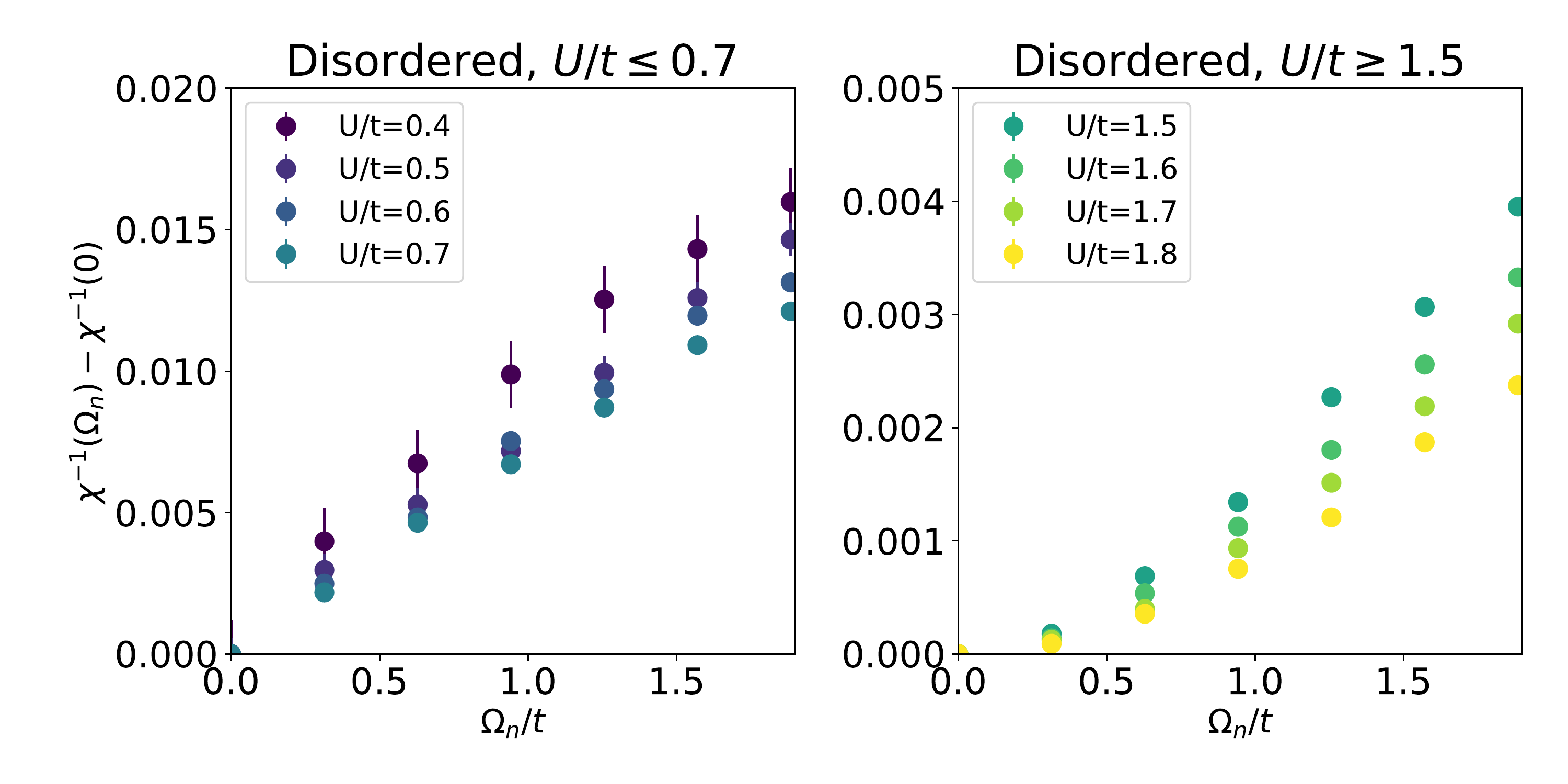}
\caption{Dynamical inverse spin susceptibility for $\beta t=20$ for $U/t \leq 0.7$ and $U/t \geq 1.5$. Despite the decreased density in points, the same trends as Figure 3 of the main text are visible. At small values of $\Omega_n$, the inverse dynamical susceptibility is linear in $\Omega_n$ on the left side of the magnetic dome, for $U/t \leq 0.7$, and nearly quadratic in frequency for $U/t \geq 1.5$.}
\label{fig:dyn_susc_beta_20}
\end{figure}
We conclude by showing the dynamical AFM susceptibility at $T =0.05t \gtrsim T_c$ in Fig.~\ref{fig:dyn_susc_beta_20}. This shows a behavior similar to Figure 3 of the main text for $T/t=0.025$, with the inverse susceptibility being linear in $\Omega_n$ on the left side of the phase diagram and nearly quadratic on the right side of the phase diagram.

\section{measuring the charge compressibility}
To probe the possibility of the appearance of an insulating phase, we measure the compressibility. It is obtained from the uniform component of the charge susceptibility, given by
\begin{equation}
    \chi_c = \frac{\beta}{L^2N_\tau}\sum_{\mathbf{r}_i,\mathbf{r}_j,\tau} \langle \delta \rho(\mathbf{r}_i,\tau)\delta\rho(\mathbf{r}_j,0)\rangle,
\end{equation}
where $\delta \rho(\mathbf{r}_i)=\sum_{\alpha}\left[\left(c^{\dagger}_{\mathbf{r}_i,\alpha}c_{\mathbf{r}_i,\alpha}+d^{\dagger}_{\mathbf{r}_i,\alpha}d_{\mathbf{r}_i,\alpha}\right) -\langle c^{\dagger}_{\mathbf{r}_i,\alpha}c_{\mathbf{r}_i,\alpha}+d^{\dagger}_{\mathbf{r}_i,\alpha}d_{\mathbf{r}_i,\alpha}\rangle \right]$. Errors in this quantity are estimated from a jackknife analysis~\cite{Gubernatis2016}. In Fig.~\ref{fig:compressibility} we show the compressibility $\frac{\mathrm{d}n}{\mathrm{d}\mu}=\chi_c$ as a function of $U$ for various system sizes and temperatures. The $L=12$ curve corresponds to the one shown in the main text. While at low temperatures the compressibility has a sharp suppression, at high temperatures the compressibility decreases smoothly towards zero. In the main text, this quantity yields the (logarithmic) color scale in Fig.~1{\sffamily{\bfseries C}} of the main text along with the black dashed line which indicates the $\chi_c=0.01$ threshold.
\begin{figure}
\includegraphics[width=0.3\textwidth]{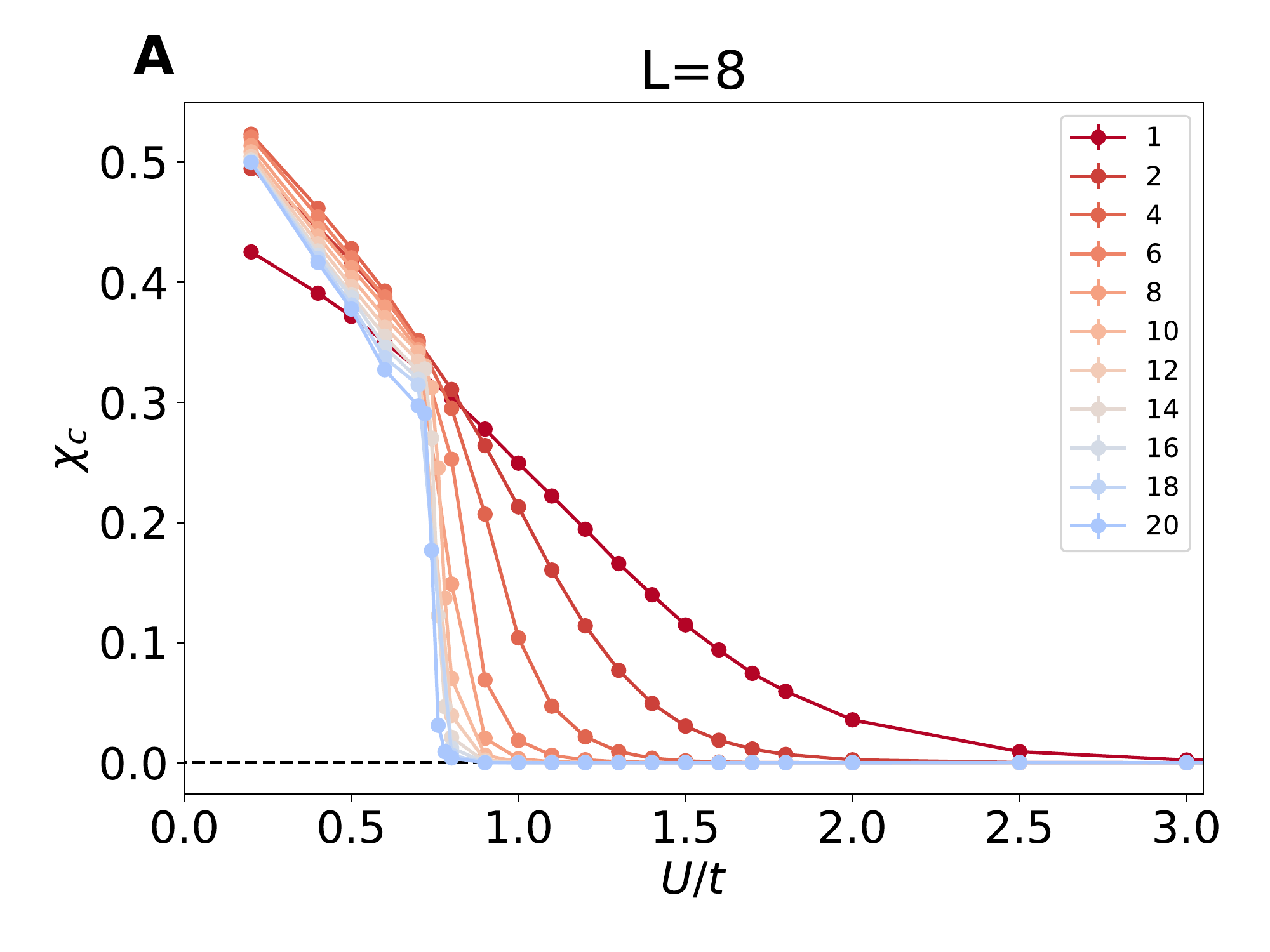}\includegraphics[width=0.3\textwidth]{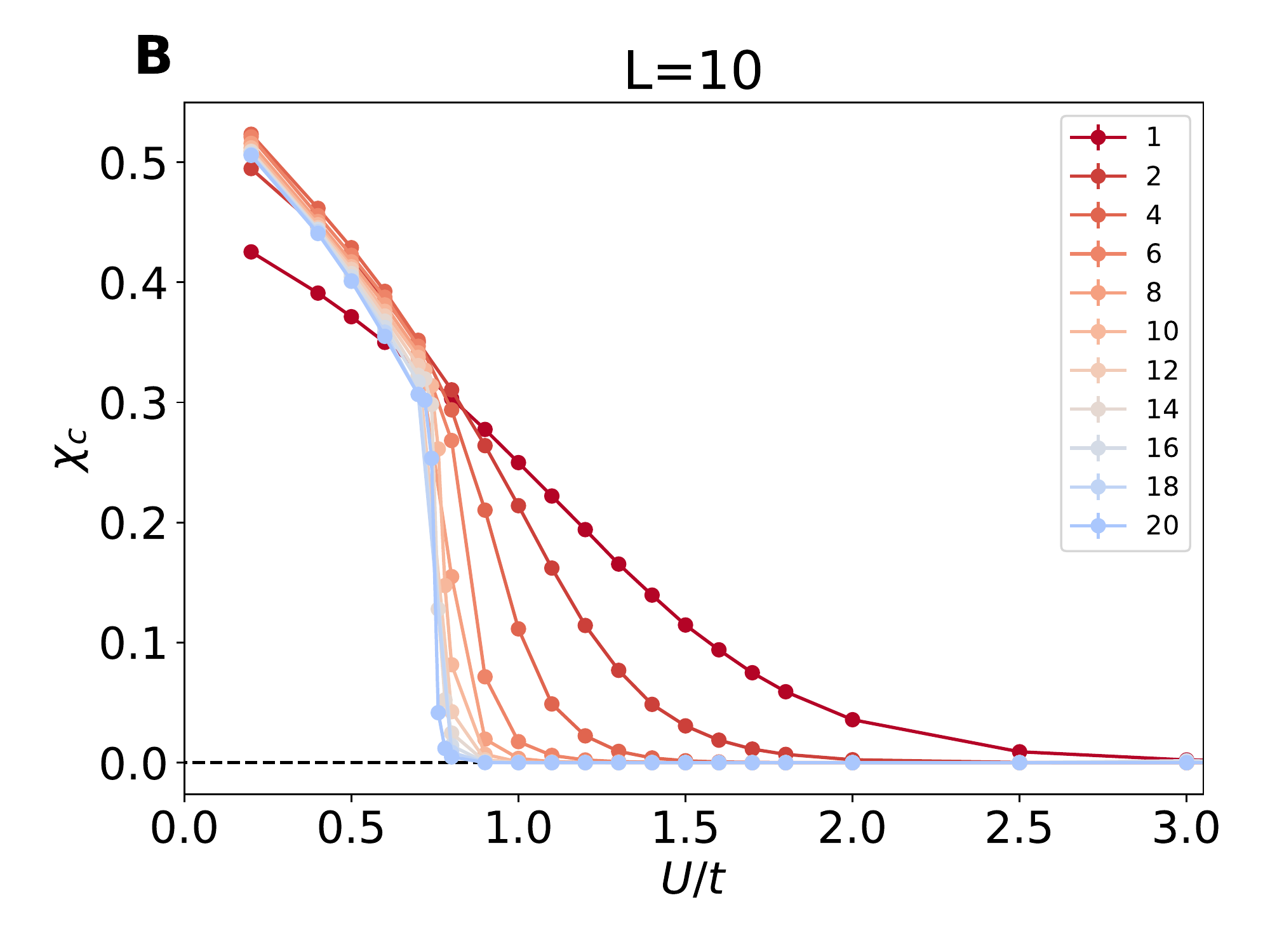}\includegraphics[width=0.3\textwidth]{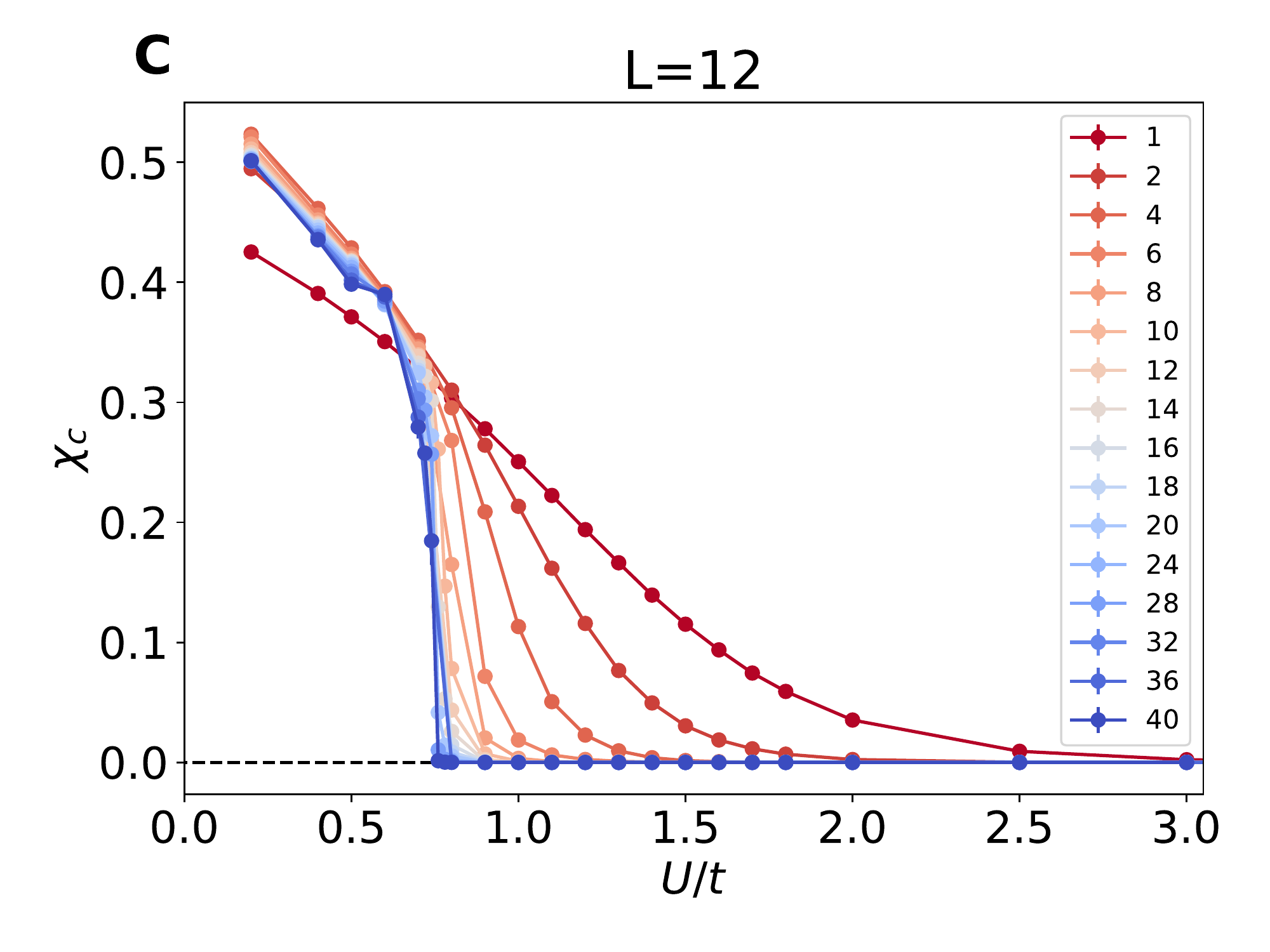}
\includegraphics[width=0.3\textwidth]{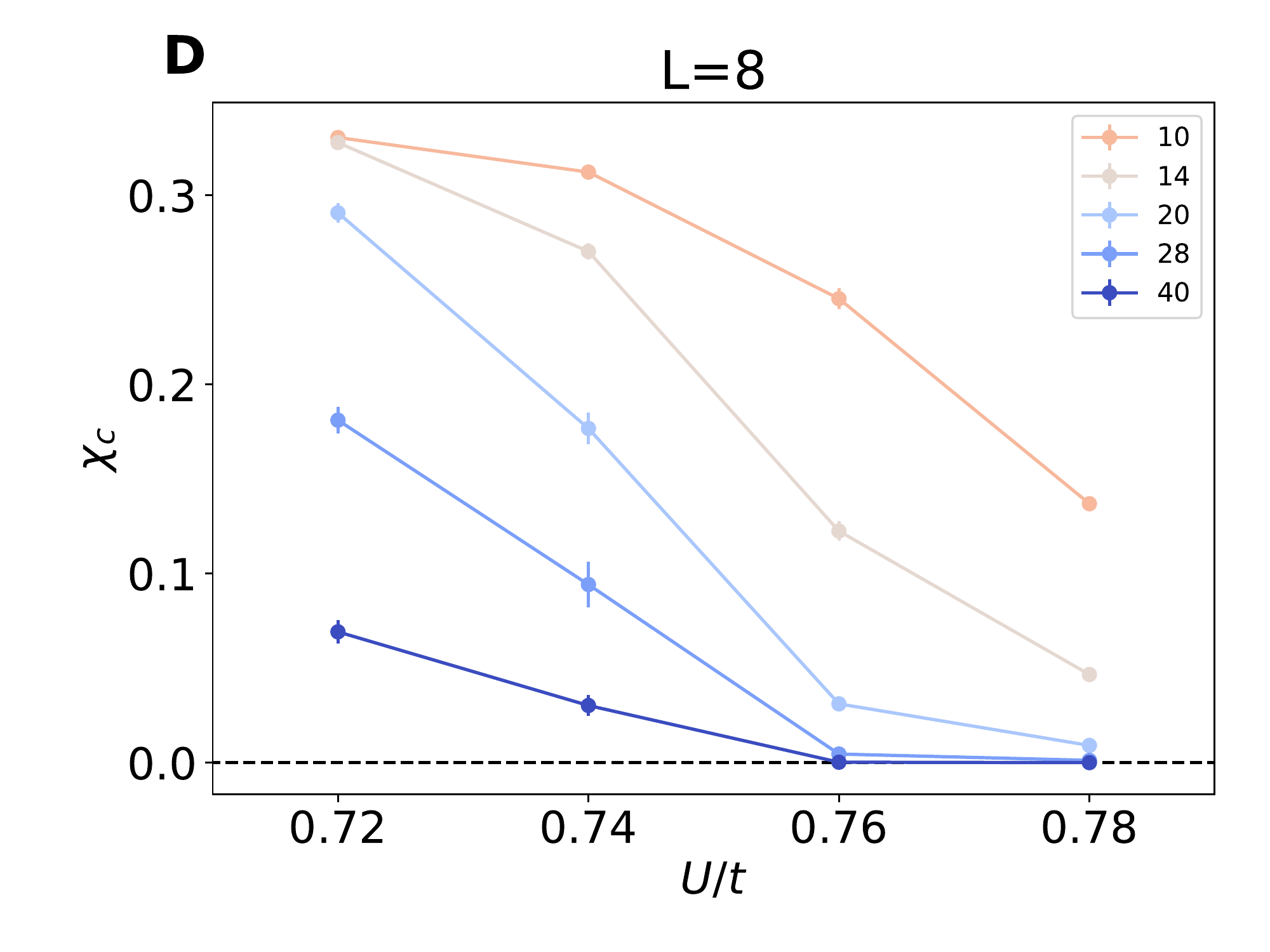}\includegraphics[width=0.3\textwidth]{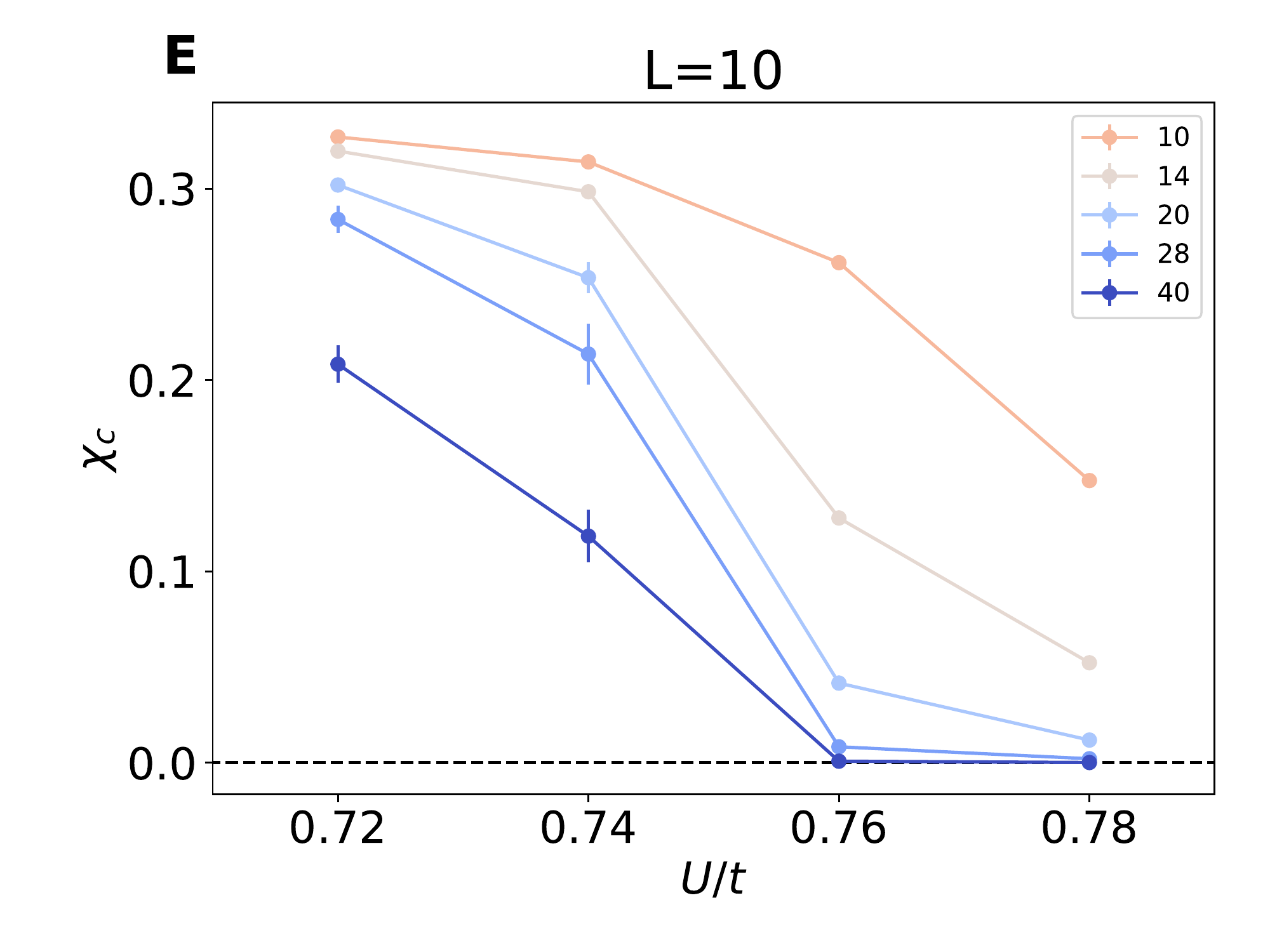}\includegraphics[width=0.3\textwidth]{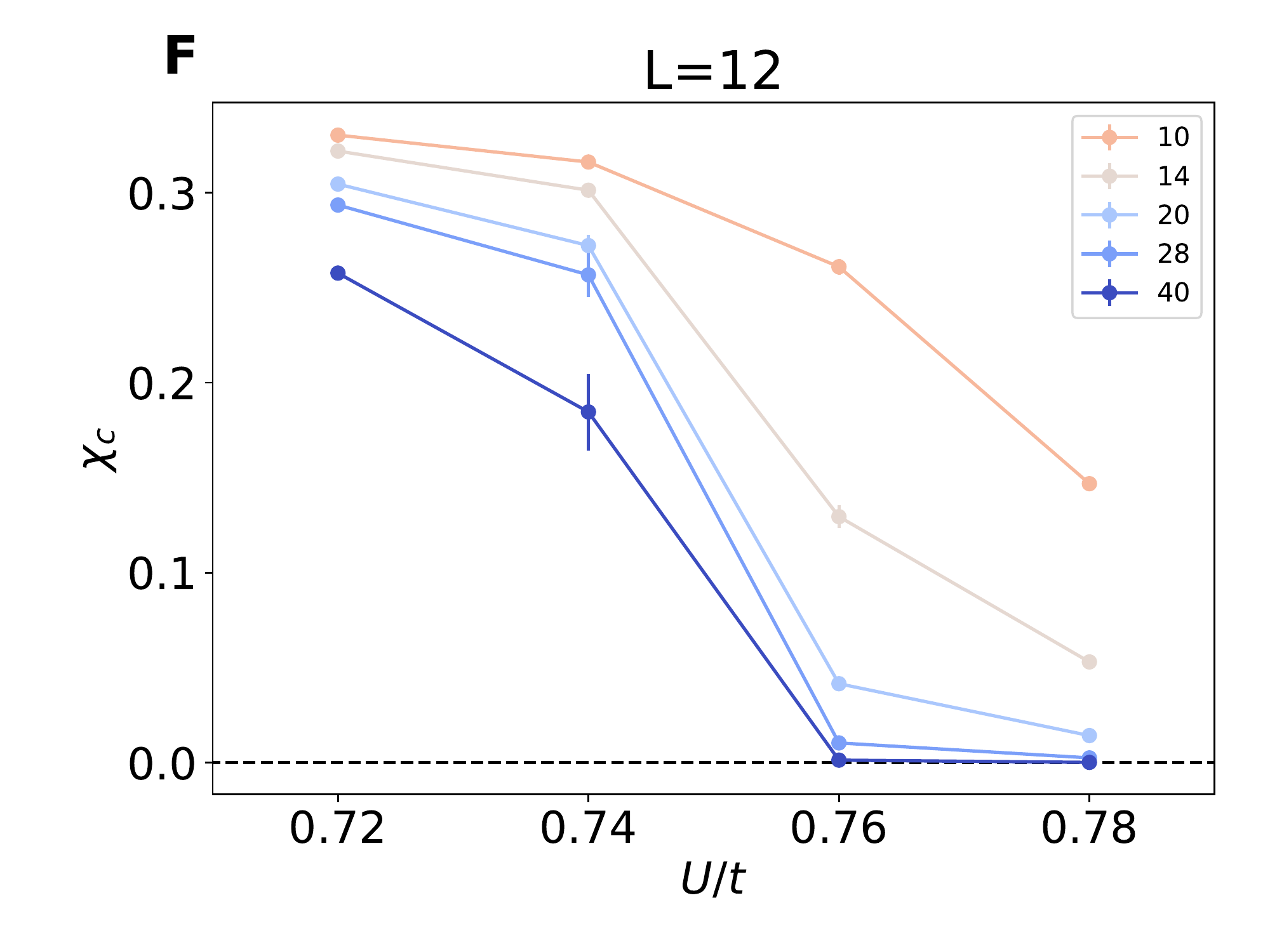}
\caption{Charge compressibility $\chi_c$ as a function of $U$ for different system sizes and values of $\beta t$ (as indicated in the legends). For all system sizes, a noticeable suppression of $\chi_c$ occurs around $U/t \approx 0.76$ at low temperatures}
\label{fig:compressibility}
\end{figure}

\section{Identifying the superconducting transition}
To study superconductivity we measure both the pair susceptibility $\chi_p$ and the superfluid density $\rho_S$. The pair susceptibility is defined in the $s_{\pm}$ channel, corresponding to a gap function that has opposite signs in the two bands. Denoting $P_{\pm}(\mathbf{r}_i)=\sum_{\alpha\beta}i\sigma_{\alpha\beta}^{y}\left(c_{i\alpha}c_{i\beta} - d_{i\alpha}d_{i\beta}\right)$, we have:
\begin{equation}
    \chi_p = \frac{\beta}{L^2N_\tau}\sum_{\mathbf{r}_i,\mathbf{r}_j,\tau} \langle P^\dagger_{\pm}(\mathbf{r}_i,\tau)P_{\pm}(\mathbf{r}_j,0)\rangle. 
\end{equation}
This quantity is shown in Fig.~\ref{fig:pair_susc} for different system sizes. As also shown in the main text, this quantity displays a suppression as we enter the magnetic phase and no subsequent increase once the magnetic order subsides.
\begin{figure}
\includegraphics[width=0.3\textwidth]{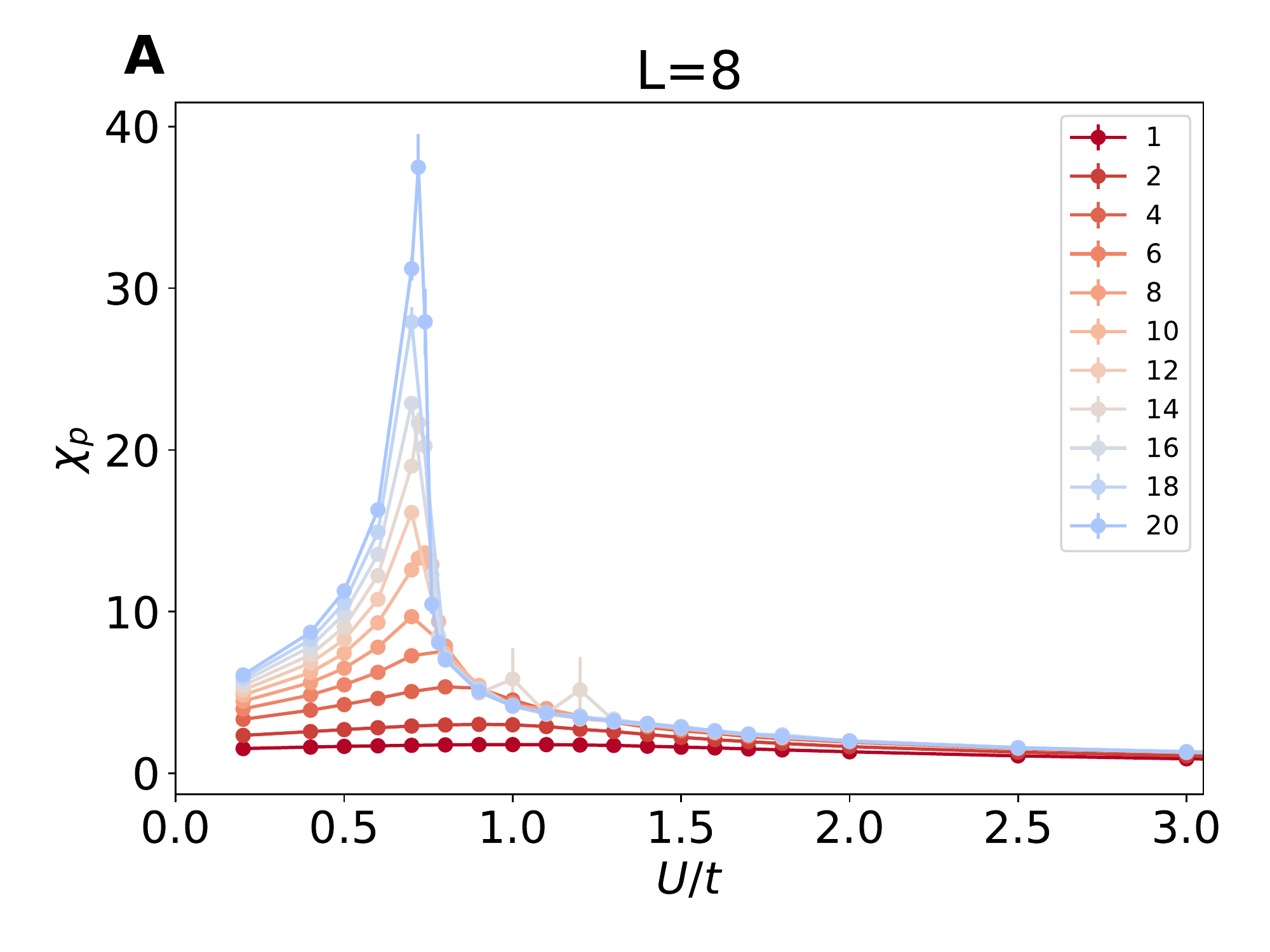}\includegraphics[width=0.3\textwidth]{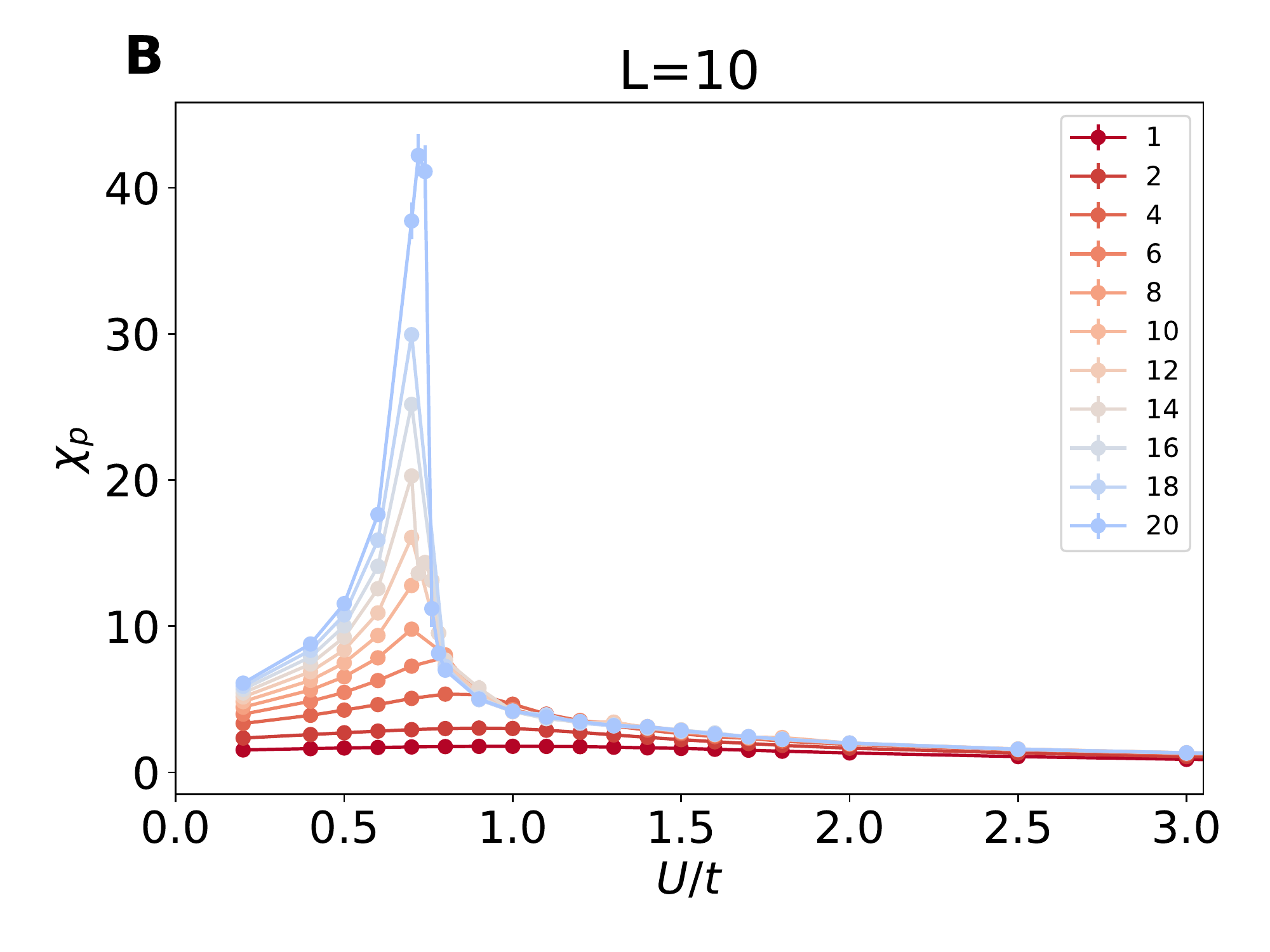}\includegraphics[width=0.3\textwidth]{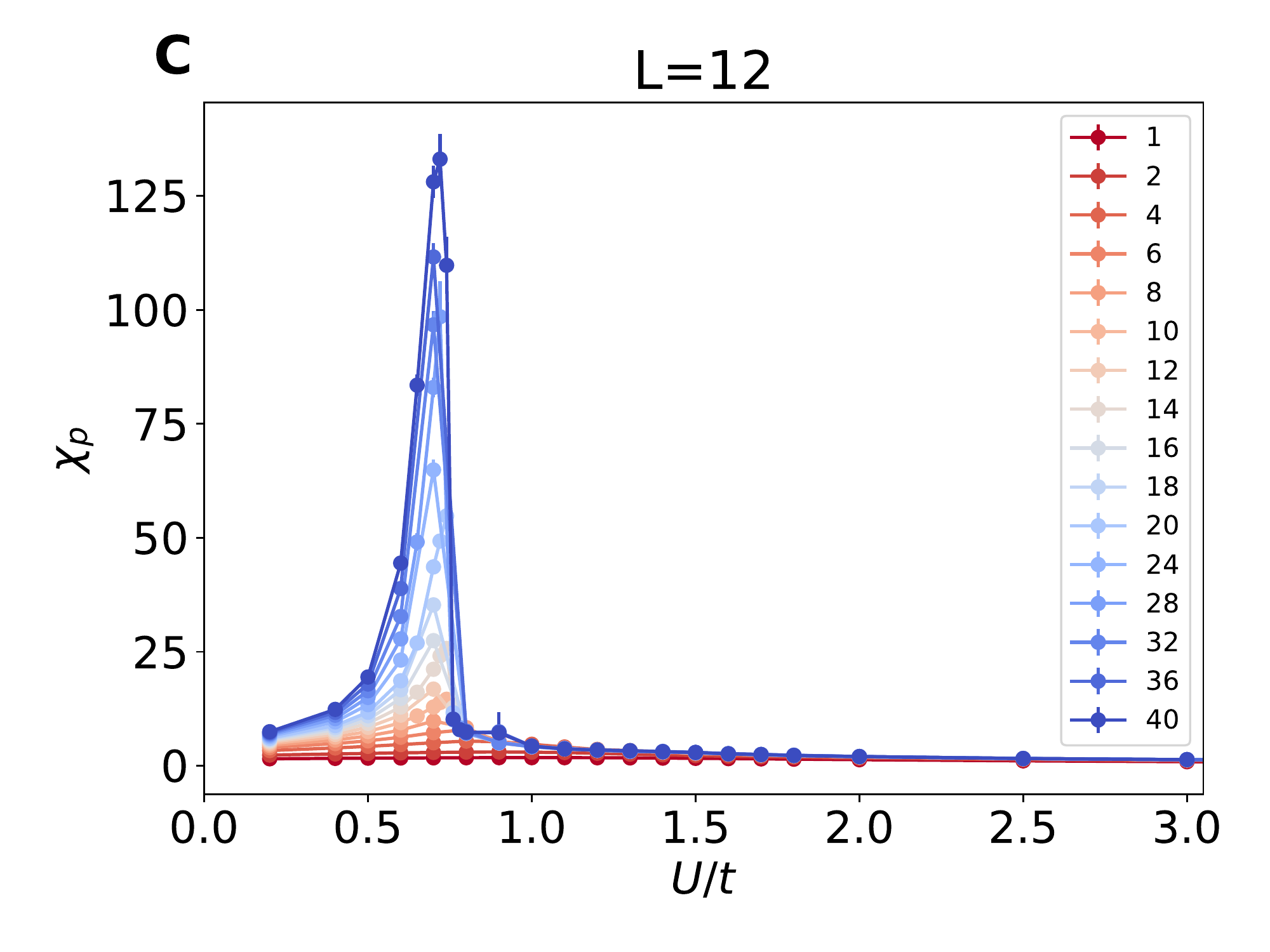}
\includegraphics[width=0.3\textwidth]{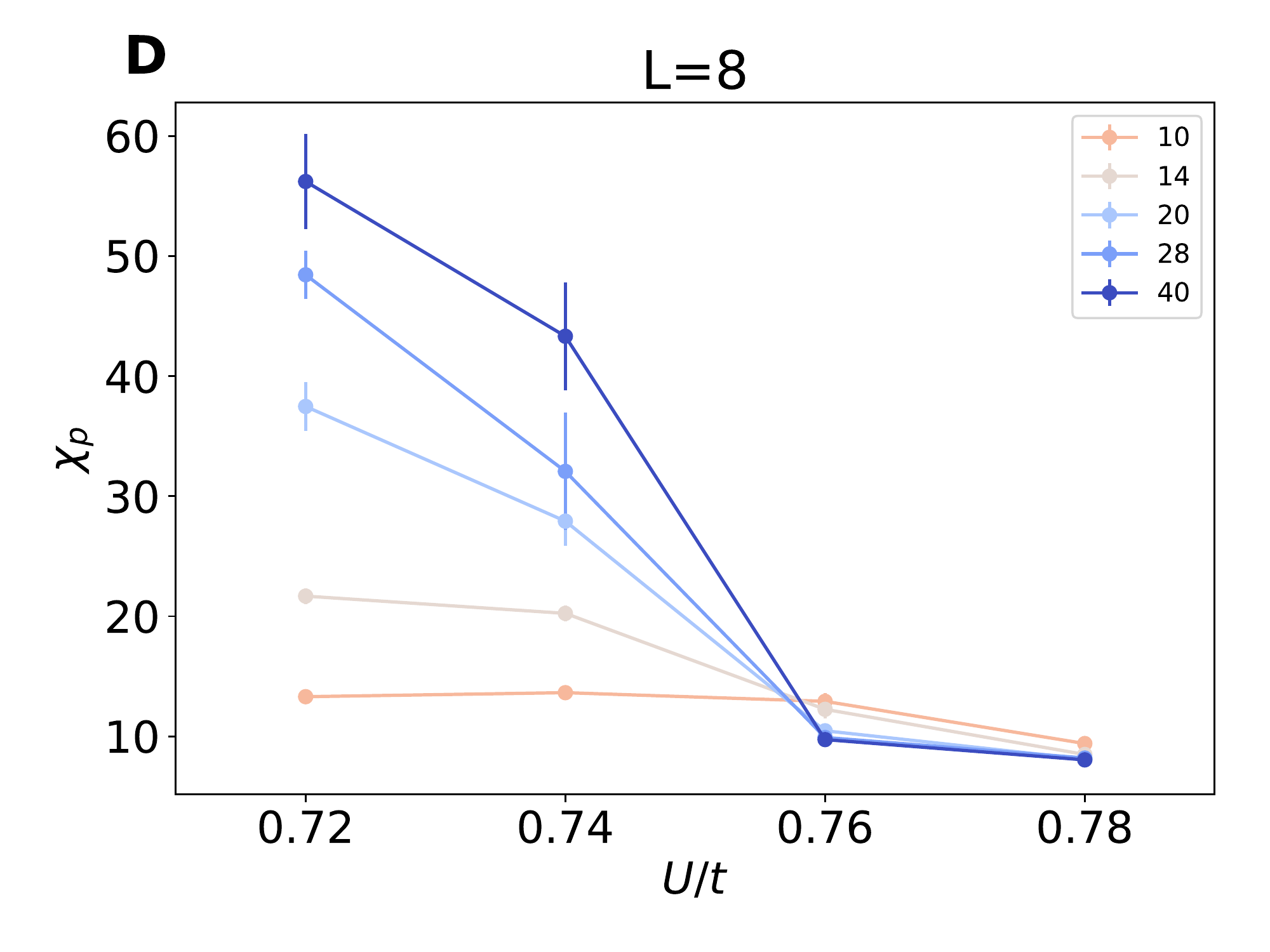}\includegraphics[width=0.3\textwidth]{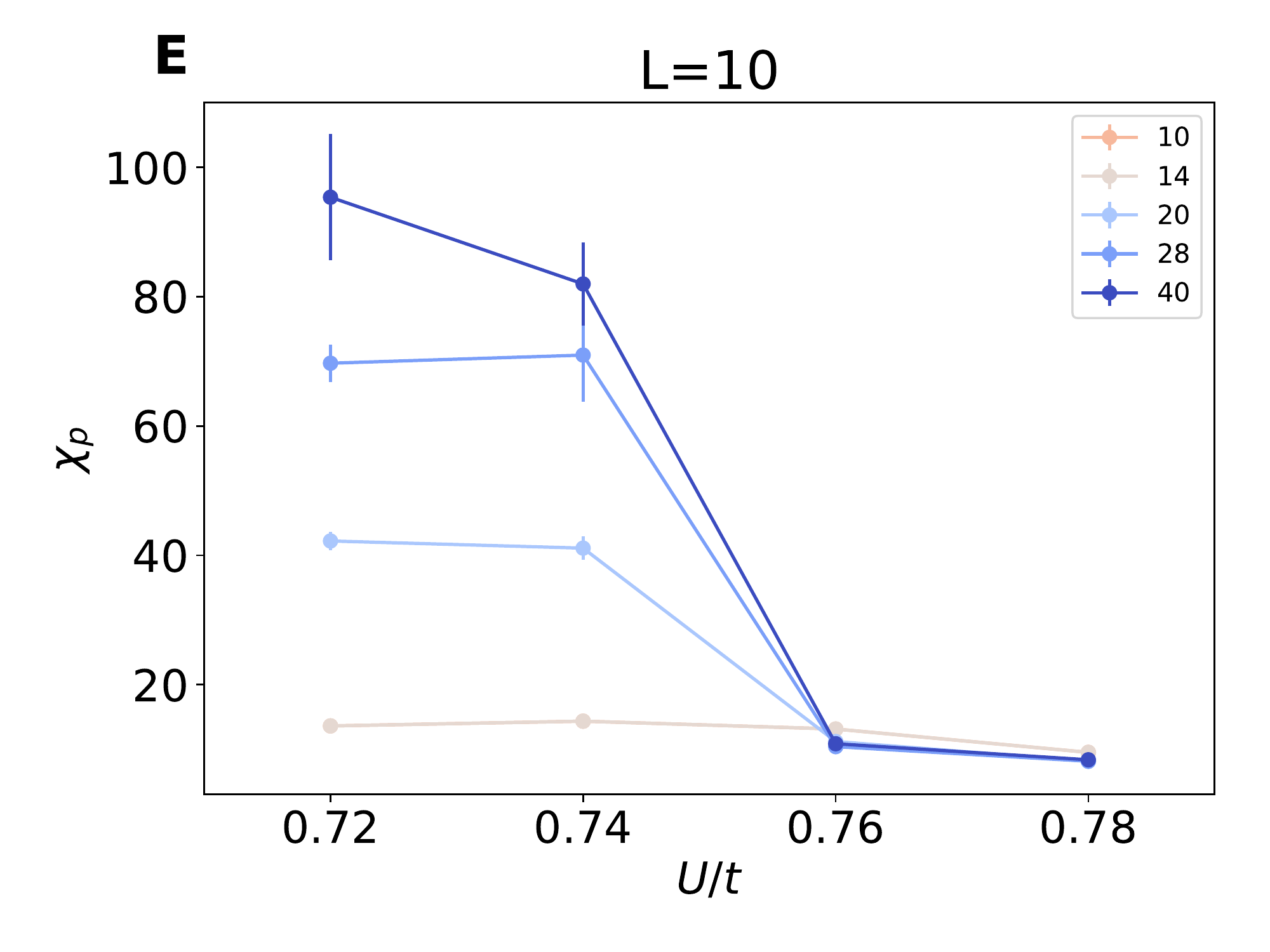}\includegraphics[width=0.3\textwidth]{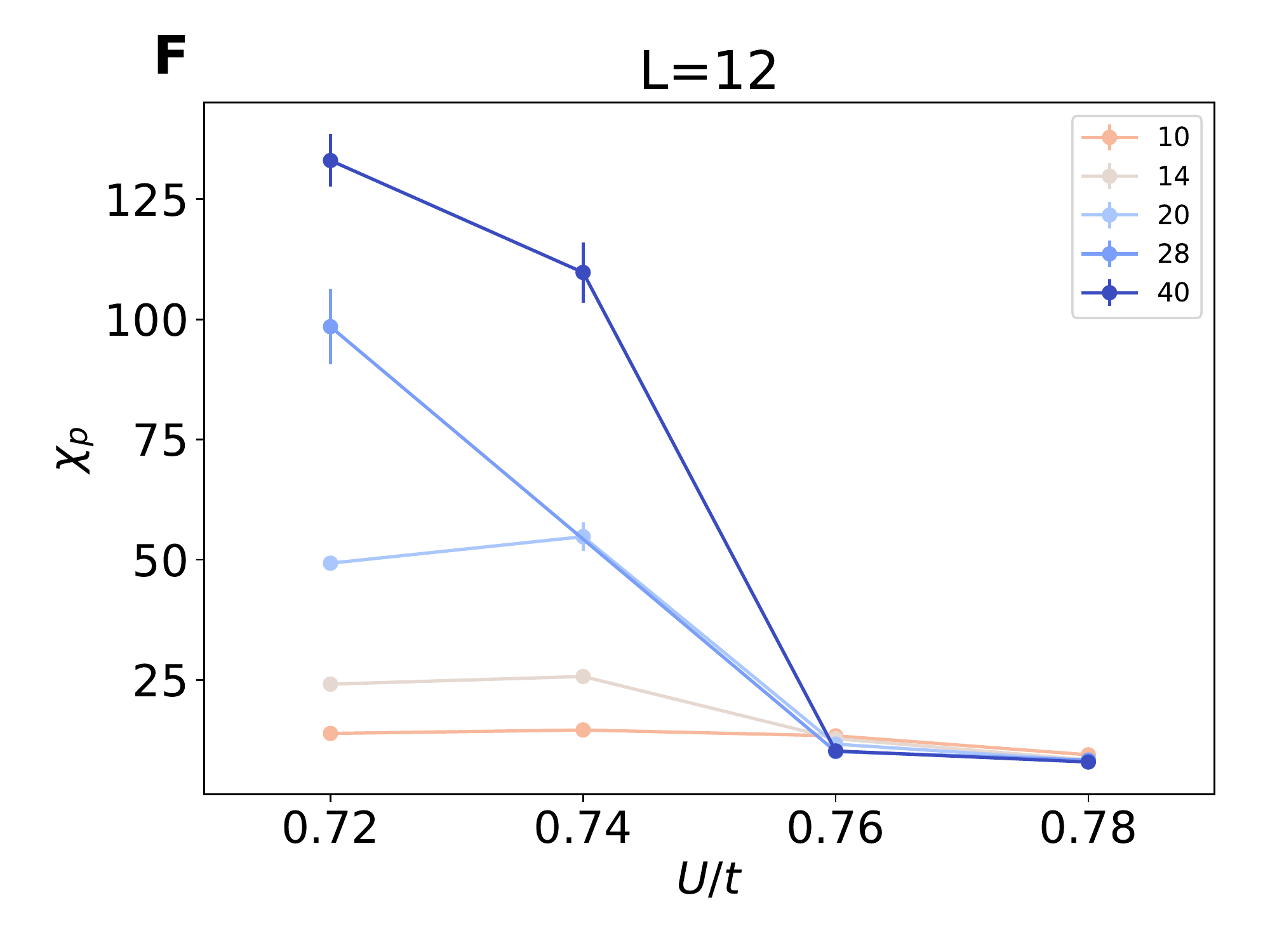}
\caption{Pair susceptibility $\chi_p$ as a function of $U$ for different system sizes and values of $\beta t$ (as indicated in the legends). Regardless of system size, there is a rapid suppression near $U/t=0.76$, where the system develops long-range magnetic order and the compressibility is sharply depleted.}
\label{fig:pair_susc}
\end{figure}

The superfluid density is a thermodynamic measure for a superconducting state regardless of the pairing form factor, and is defined as \cite{scalapino93}:
\begin{equation}
    \rho_s = \frac{1}{4}\left[\Lambda_{xx}(q_x\rightarrow 0,q_y=0,i\Omega_n=0)-\Lambda_{xx}(q_x=0,q_y\rightarrow 0,i\Omega_n=0)\right],
\end{equation}
where
\begin{equation}
    \Lambda_{xx}(\mathbf{r}_i,\tau) = \frac{1}{ L^2}\sum_{\mathbf{r}_j}\langle J_x(\mathbf{r}_i+\mathbf{r}_j,\tau)J_x(\mathbf{r}_j,0)\rangle
\end{equation}
is the current-current correlation function.
The superfluid density for different system sizes and temperatures is shown in Fig.~\ref{fig:sup_dens}. The drastic suppression upon entering the magnetic phase is evident for all values of $L$. Using the Berezinskii-Kosterlitz-Thouless (BKT) criterion, $\rho_s(T_c) = \frac{2T_c}{\pi}$, we determine $T_c$. In two dimensions, the BKT temperature scales logarithmically with system size \cite{sandvik2011}, and no scaling analysis was carried out for this quantity.
\begin{figure}
\includegraphics[width=0.3\textwidth]{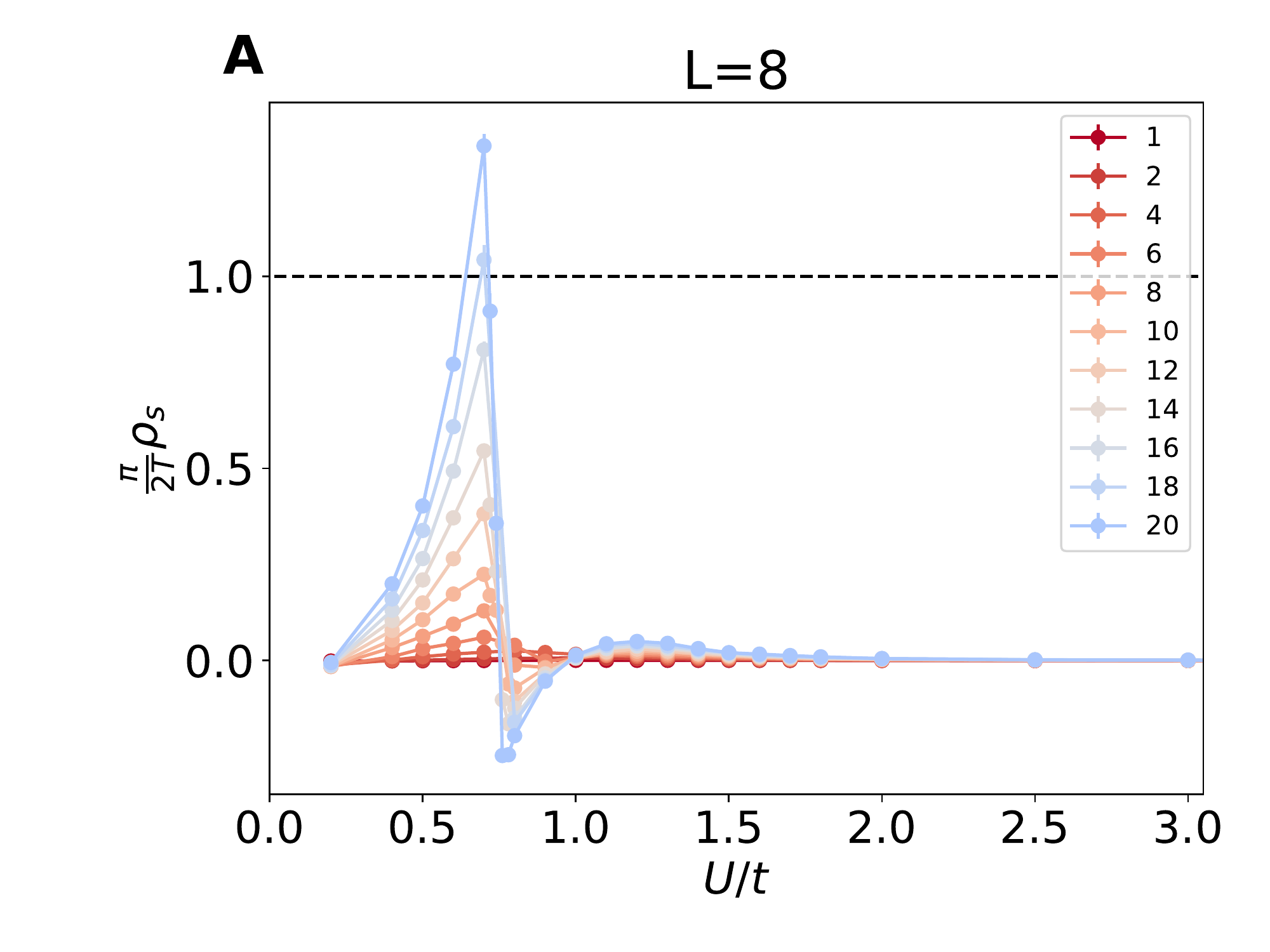}\includegraphics[width=0.3\textwidth]{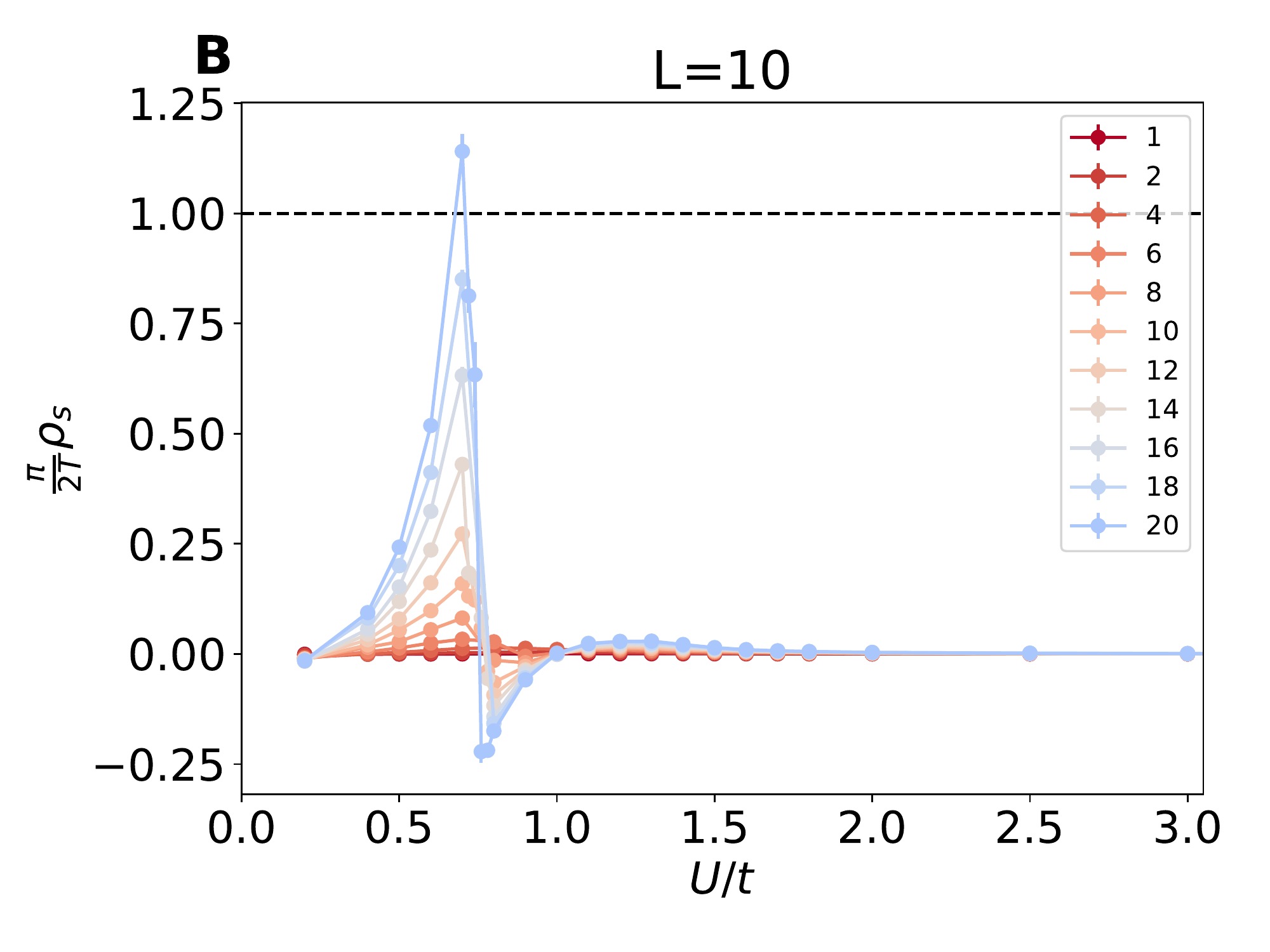}\includegraphics[width=0.3\textwidth]{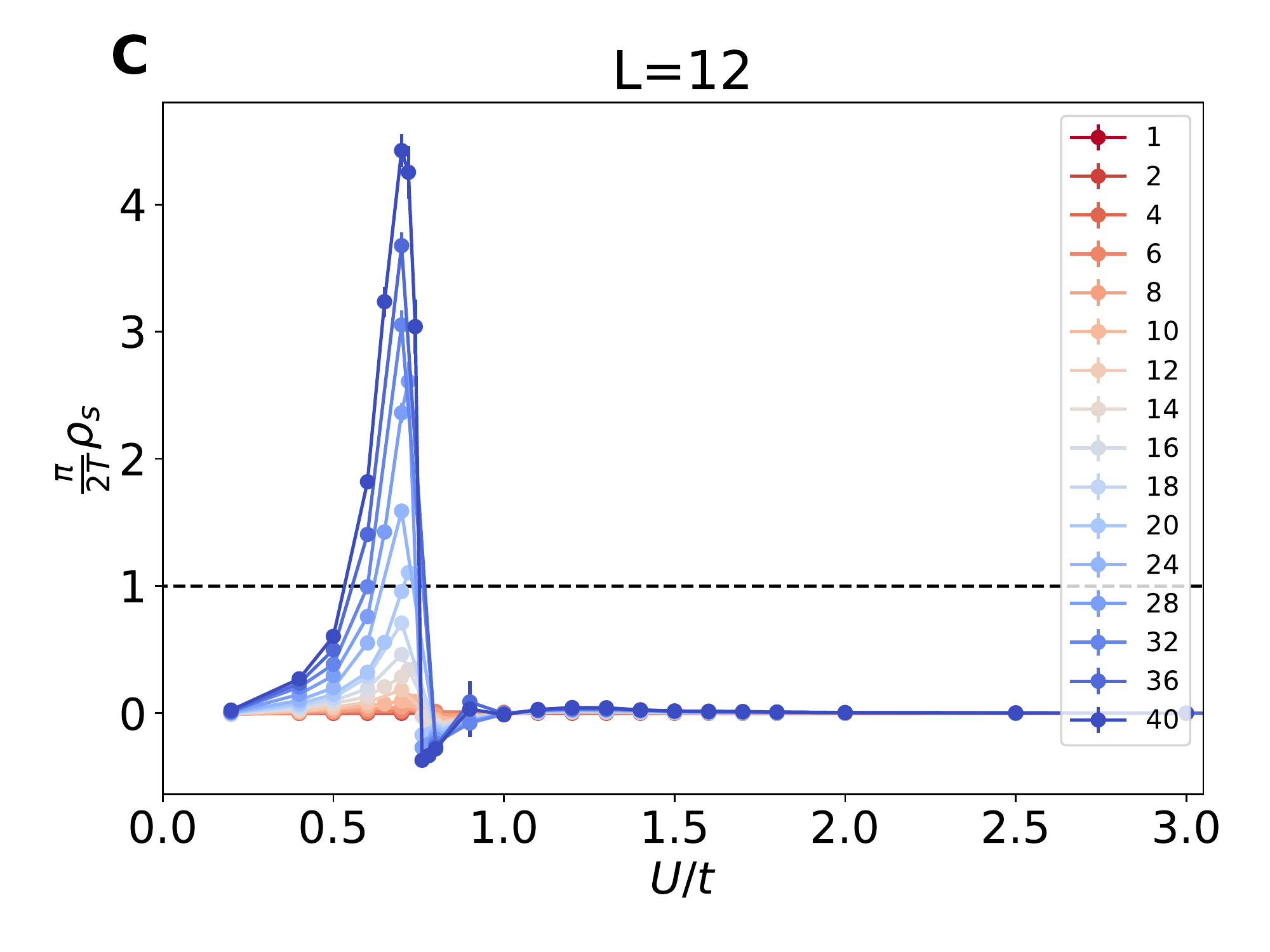}
\includegraphics[width=0.3\textwidth]{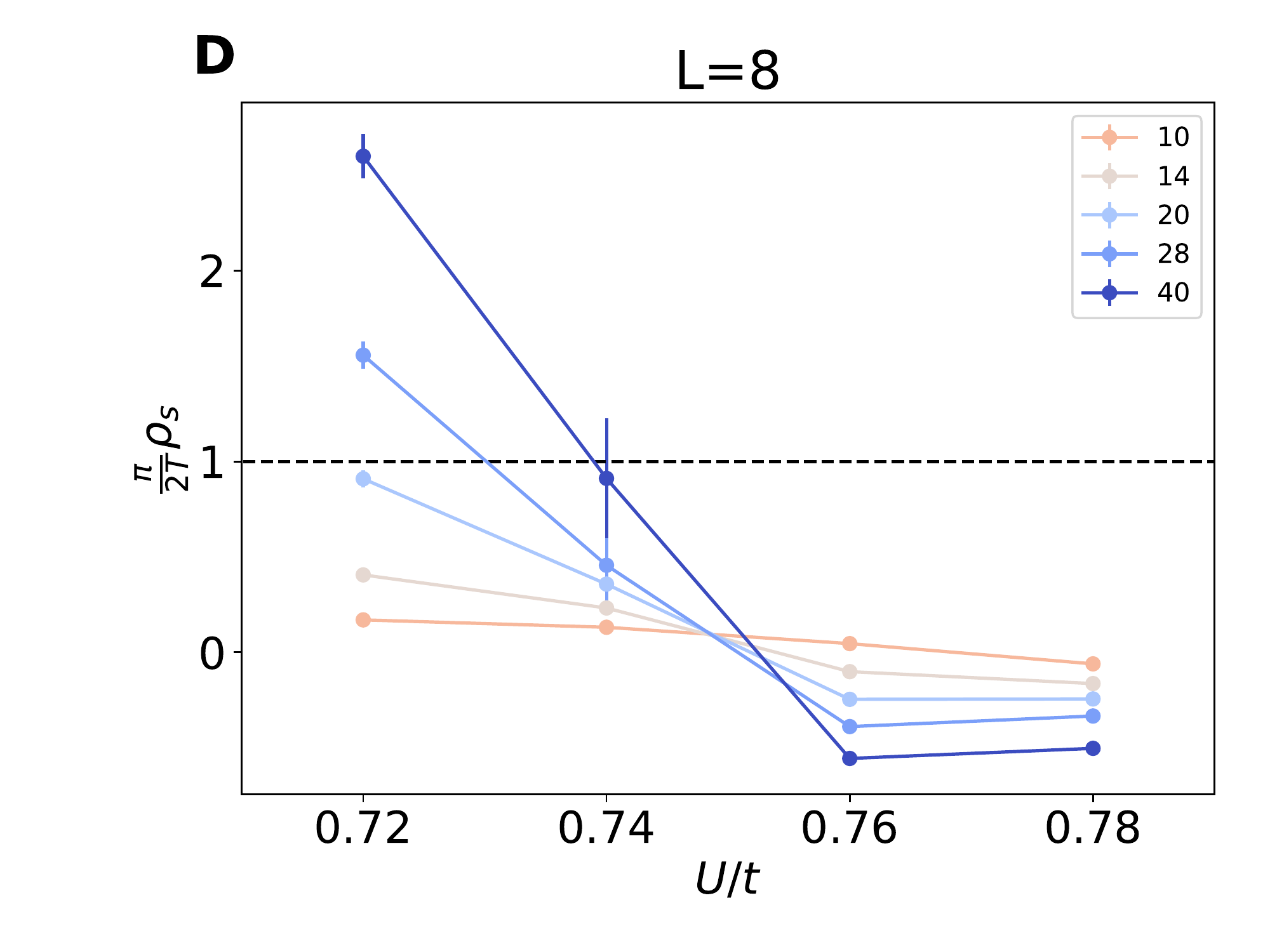}\includegraphics[width=0.3\textwidth]{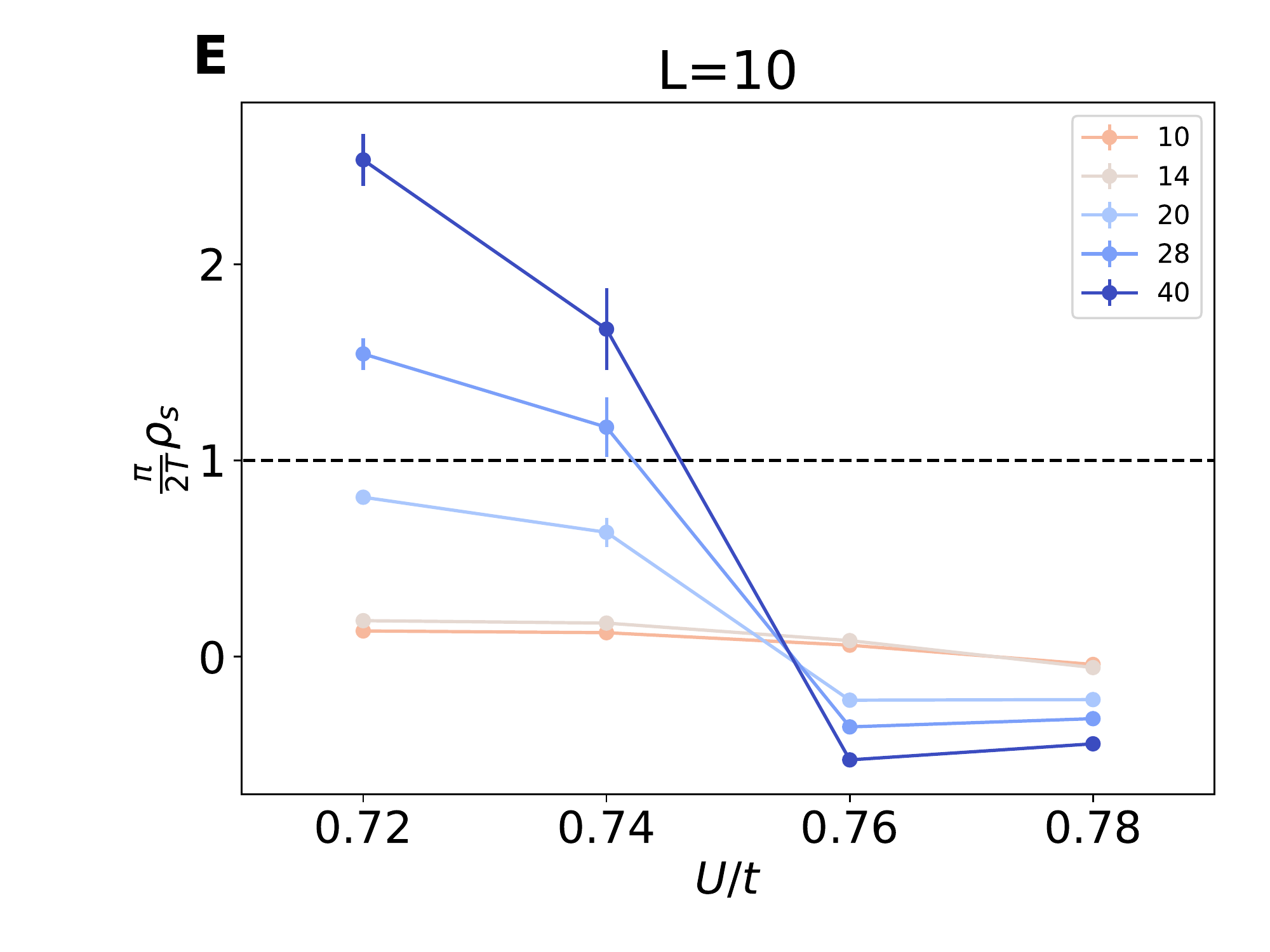}\includegraphics[width=0.3\textwidth]{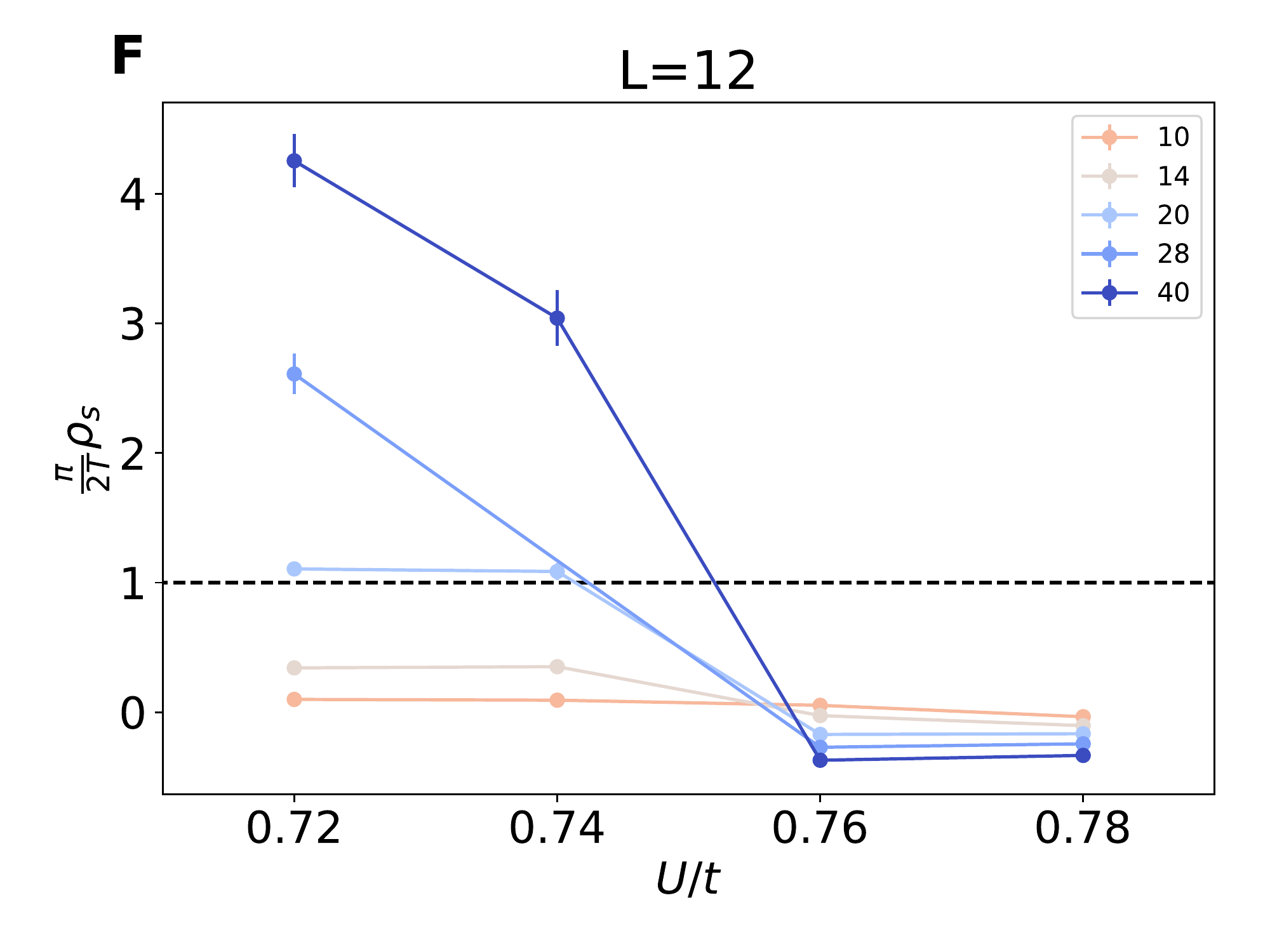}
\caption{Superfluid density $\rho_s$ as a function of $U$ for different system sizes and values of $\beta t$ (as indicated in the legends). The dashed line denotes the BKT criterion. Note the massive suppression associated with the onset of magnetic order beyond $U/t=0.76$. This quantity was not measured for $L=14$.}
\label{fig:sup_dens}
\end{figure}

\section{Finite size effects in the spectral weight proxy $\tilde{Z}_{\mathbf{k}}$}

To elucidate the possible effects of finite-system sizes on the spectral weight proxy, $\tilde{Z}_{\mathbf{k}}$, plotted in Fig.~3 of the main text, we here include data for the $L=8$ and $L=10$ cases as well. As in the main text, we have combined simulations from 16 different twisted boundary conditions, which serves to alleviate effects of finite-system sizes. In Fig.~\ref{fig:fs_system_sizes} we show $\tilde{Z}_{\mathbf{k}}$ for $U/t=0.8$ for $L=8$, $L=10$, and $L=12$. Note that in the main text the color bar was rounded off at the second digit.
\begin{figure}
\includegraphics[width=0.7\textwidth]{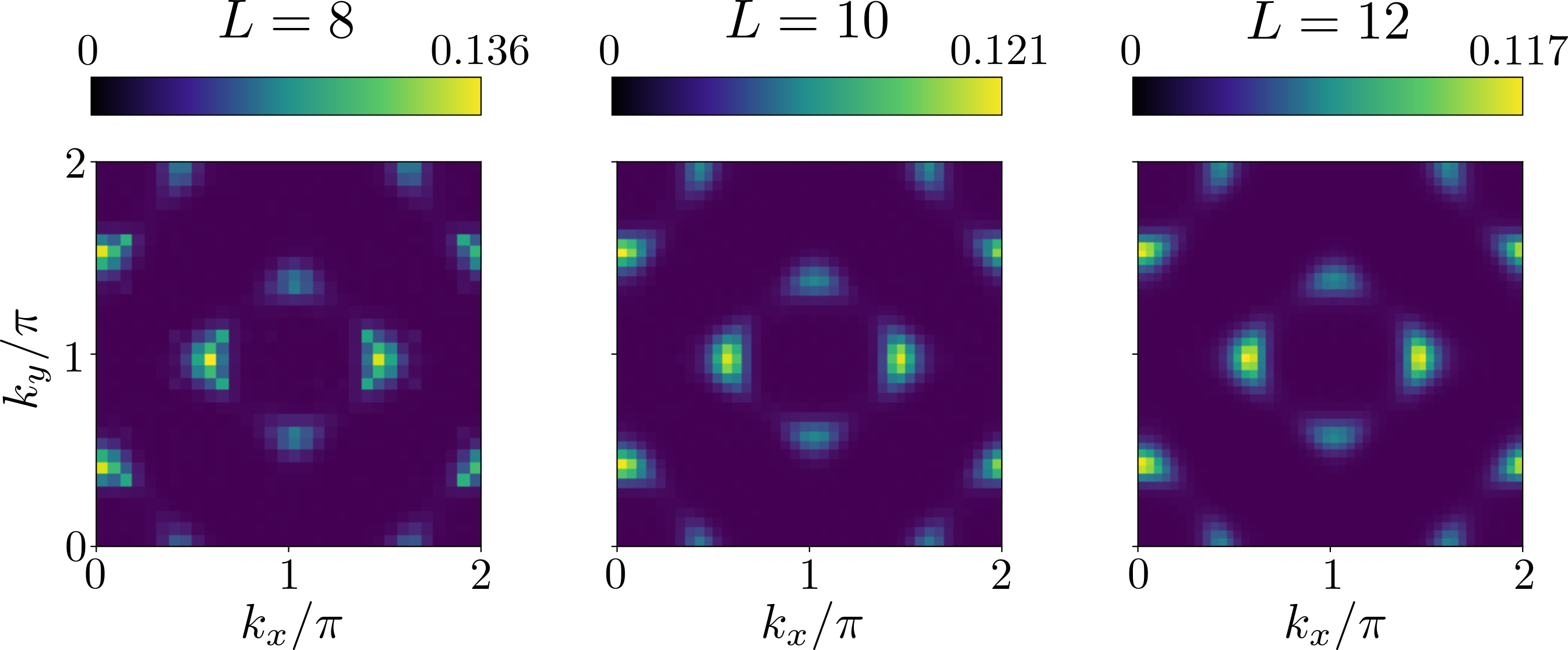}
\caption{\label{fig:fs_system_sizes}Comparison of the spectral weight proxy, $\tilde{Z}_{\mathbf{k}}$, for different system sizes. Here, as in the main text, we have combined simulations from 16 different twisted boundary conditions to alleviate finite-size effects.}
\end{figure}
Furthermore, in Fig.~\ref{fig:fs_cut_as_L}, we show cuts through $\tilde{Z}_{\mathbf{k}}$ for $k_y=\pi$ and $k_x=\pi$, respectively. The fact that the points from different system sizes fall on the same line gives us confidence that finite-size effects are minor.
\begin{figure}
\includegraphics[width=0.4\textwidth]{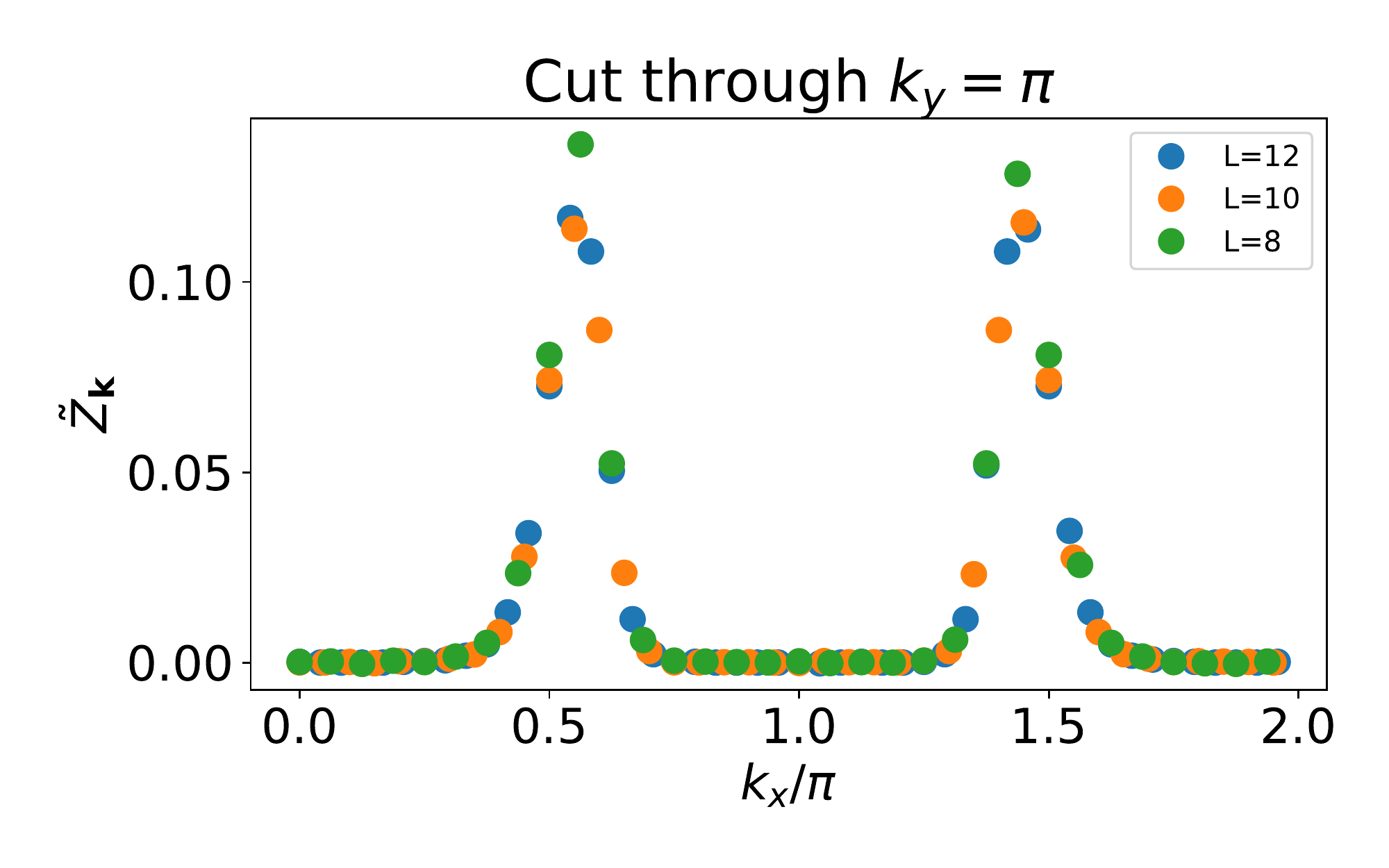}\includegraphics[width=0.4\textwidth]{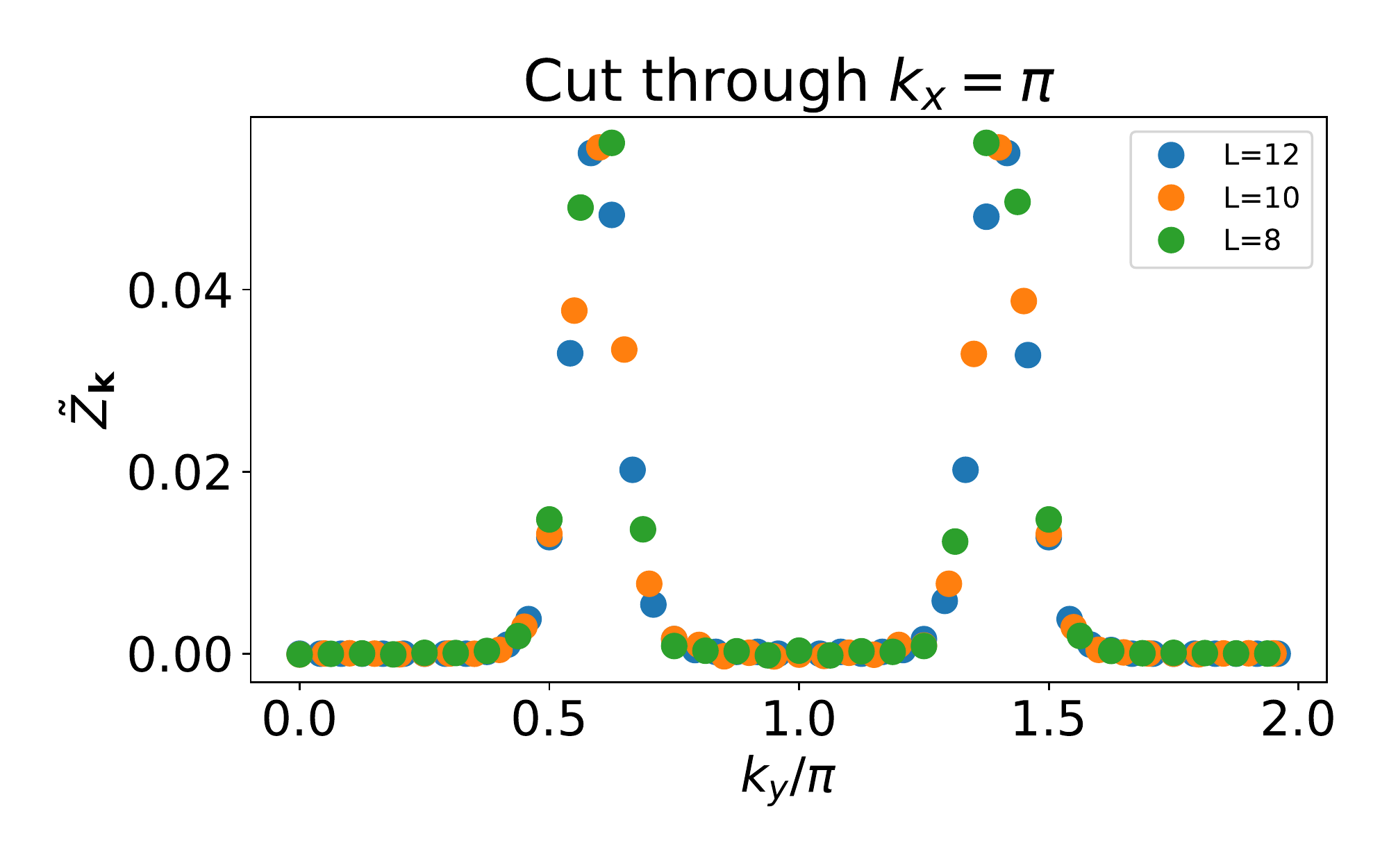}
\caption{\label{fig:fs_cut_as_L} Cuts through $\tilde{Z}_{\mathbf{k}}$ shown in Fig.~\ref{fig:fs_system_sizes} for $k_y=\pi$ and $k_x=\pi$, respectively. The fact that these points all fall on the same line suggests that finite-size effects are minor.}
\end{figure}

\section{Strong coupling expansion}
Here we show that in the strong coupling limit, where $U\gg  t, \, \delta, \, \mu$, our two-band model maps onto a transverse-field Ising model in two spatial dimensions. Depending on the ratio $\mu^2/(t^2-\delta^2)$, the system exhibits a quantum phase transition between an Ising antiferromagnet and a quantum paramagnet, where the local spinor is composed of linear superposition of the spin and band degrees of freedom. 

To zeroth order in the kinetic energy terms, it suffices to study a single site. We define two fermionic annihilation operators as $\gamma_{1\br\alpha} = \frac{1}{\sqrt{2}}\left( c_{\br\alpha} + d_{\br\alpha}\right)$ and $\gamma_{2\br\alpha} = \frac{1}{\sqrt{2}}\left( c_{\br\alpha} - d_{\br\alpha}\right)$, corresponding to ``bonding" and ``anti-bonding" combinations of the two bands. In the new basis, the interaction term can be written as:
\begin{equation}
    H_U = - U \sum_\mathbf{r} \left( \gamma^{\dagger}_{1\mathbf{r}\alpha}\sigma^z_{\alpha\beta}\gamma_{1\mathbf{r}\beta}-\gamma^{\dagger}_{2\mathbf{r}\alpha}\sigma^z_{\alpha\beta}\gamma_{2\mathbf{r}\beta}\right)^2.
\end{equation}
The energy is minimized when the bonding and anti-bonding states are both polarized but with opposite spins, yielding the ground state energy $E_U = -4U L^2$, where $L$ is the linear dimension of the system. Therefore, this ground state is degenerate and possesses a local SU(2) symmetry corresponding to a combined rotation in spin and bonding space. It can be written as 
\begin{equation}
    |\Psi_{g.s.}\rangle = \Pi_{\mathbf{r}} \left( u_{\mathbf{r}}|a_\mathbf{r}\rangle + v_{\mathbf{r}}|b_\mathbf{r}\rangle \right) 
\end{equation}
where $|u_{\mathbf{r}}|^2+|v_{\mathbf{r}}|^2=1$, and we have defined the two basis states: $|a_\mathbf{r}\rangle \equiv \gamma^{\dagger}_{1\mathbf{r}\uparrow}\gamma^{\dagger}_{2\mathbf{r}\downarrow} |0\rangle,\ |b_\mathbf{r}\rangle= \gamma^{\dagger}_{1\mathbf{r}\downarrow}\gamma^{\dagger}_{2\mathbf{r}\uparrow}|0\rangle $. 

Next we perform second order perturbation theory in the kinetic term, and work in the projected Hilbert space. We find:
\begin{equation}
    \mathcal{H} = -\left[\sum_{\mathbf{r}} 
    \left( 4U s^0 + \frac{\mu^2}{4U} \right) 
   - \frac{t^2+\delta^2}{3U}\sum_{\mathbf{r}\mathbf{r}'} s^0_{\mathbf{r}}s^0_{\mathbf{r}'} \right] +\left[ \frac{\mu^2}{4U}\sum_{\mathbf{r}} s_{\mathbf{r}}^x
   + \frac{t^2-\delta^2}{3U}\sum_{
   \langle \mathbf{r}\mathbf{r}' \rangle} s^z_{\mathbf{r}}s^z_{\mathbf{r}'} \right].
\end{equation}
Here $(s^0,\vec{s})$ are the identity and Pauli matrices acting on the projected Hilbert space $|\psi_\mathbf{r}\rangle \equiv (|a_\mathbf{r}\rangle,|b_\mathbf{r}\rangle)$. The first term are constants, which reduce to the ground state energy $-4UL^2$ in the infinite $U$ limit. The second term corresponds to a transverse-field Ising model with a transverse field $h=\mu^2/4U$ and anti-ferromagnetic exchange interaction $J=(t^2-\delta^2)/3U$. 

The transverse field Ising model has a quantum phase transition at $h_c/J\approx 3$~\cite{Jongh1998}. For the band parameters in the paper, we have $h/J \approx 3.57$, meaning that in the strong coupling limit, the system is a featureless quantum paramagnetic insulator. This implies the existence of a quantum phase transition at intermediate coupling, since at weak-coupling a magnetic ordered phase exists.

\end{widetext}

\end{document}